# Modulation of structural short-range order due to chemical patterning in multi-component amorphous interfacial complexions

Esther C. Hessong [1,5], Zhengyu Zhang [2,5], Tianjiao Lei [1], Mingjie Xu [3], Toshihiro Aoki [3], Timothy J. Rupert [1,2,4,*]

[1] Department of Materials Science and Engineering, University of California, Irvine, CA 92697, USA

[2] Hopkins Extreme Materials Institute, Johns Hopkins University, Baltimore, MD 21218, USA

[3] Irvine Materials Research Institute, University of California, Irvine, CA 92697, USA

[4] Department of Materials Science and Engineering, Johns Hopkins University, Baltimore, MD 21218, USA

[5] These authors contributed equally to this work.

* Corresponding Author: tim.rupert@jhu.edu

**Abstract**

Amorphous interfacial complexions have been shown to restrict grain growth and improve damage tolerance in nanocrystalline alloys, with increased chemical complexity stabilizing the complexions themselves. Here, we investigate local chemical composition and structural short-range order in Cu-rich, multi-component nanocrystalline alloys to understand how dopants self-organize within these amorphous complexions and how local structure is altered. High resolution scanning transmission electron microscopy and elemental analysis are used to study both grain boundaries and interphase boundaries, with chemical partitioning observed for both. Notably, the amorphous-crystalline transition region is observed to be enriched in certain dopant species and depleted of others as compared to the interior of the amorphous complexions. This chemical patterning can be explained in terms of the elemental preference for ordered or disordered grain boundary environments. As only a qualitative measure of structural short-range order can be obtained with nanobeam electron diffraction for these specimens, atomistic simulations with a custom-built machine learning interatomic potential are then used to probe how dopant patterning affects local structural state. Increased grain boundary chemical complexity is found to result in a more disordered complexion structure, with segregation to the amorphous-crystalline transition regions driving changes in local structure that are sensitive to dopant ratios. As a whole, the intimate connection between local chemistry and order in amorphous interfacial complexions is demonstrated, opening the door for microstructural engineering within the amorphous complexions themselves.

## 1. Introduction

Interfacial complexions can dramatically impact the performance of materials (see, e.g., [1, 2] where grain boundary strength was altered by complexion formation), an effect which is emphasized in nanostructured materials with a high density of grain boundaries [3-5]. These interfacial states are in thermodynamic equilibrium with the surrounding grains and can be either structurally ordered or disordered [6-8]. Disordered, amorphous complexions typically have nanometer-scale thicknesses, and have been shown to improve damage tolerance, toughness, and strength in nanostructured metal alloys and ordered superlattice materials [9-11]. In nanocrystalline alloys specifically, amorphous complexions can result in improved microstructural stability [12-14], dislocation absorption [15], fracture resistance [16, 17], and radiation damage tolerance [18-20].

Interfacial segregation has been shown to be an important part of the stabilizing effect of amorphous complexions [21-23]. Multi-component segregation is particularly advantageous for further stabilizing nanocrystalline grains at very high temperatures for extended times [24] and can even stabilize the complexions themselves [21]. While the distribution of dopants has not been studied to date for amorphous complexions, prior work has shown that dopant partitioning can be important in segregated boundaries with more ordered structures. Koenig et al. [25] discovered chemical partitioning in a nanocrystalline Ni-Cu-P alloy, where P rapidly wet the grain boundaries and eventually led to grain boundary precipitation. Inhomogeneous dopant partitioning can also be influenced by grain boundary character and therefore local defect structure [1, 26]. Co-segregation of dopants with both larger and smaller size mismatch in Mg-rich multi-component alloys was reported to be more effective at stabilizing the grain structure than individual dopant segregation [27, 28]. Chemical patterning within grain boundaries can also alter important property variations, such as segregation competition between B and C in a $B_4C$-doped metastable $Fe_{40}Mn_{20}Co_{20}Cr_{15}Si_5$ high entropy alloy that was found to be critical for the identification of crack-free processing windows [29].

In addition to composition, amorphous complexions have spatial variations in structural short-range order (SSRO). Prior research has used recurring structural motifs to quantify local ordering in amorphous solids [30, 31]. The consideration of atomic arrangements within the first coordination shell represents SSRO, which has been identified as an important structural descriptor for metallic glasses. For example, icosahedral SSRO, where all first coordination shell atoms have five-fold symmetry, has been shown to influence the glass-forming ability [32, 33], transport properties [34], and mechanical properties such as strength and plastic deformation modes [35, 36]. The geometrically-favored icosahedral motif plays a key role in the deformation response of bulk amorphous metals [37, 38], while clusters of geometrically unfavored motifs also contribute to shear banding by creating soft regions prone to instability [39]. Local structural ordering can also be defined for amorphous films sandwiched between crystalline phases and is important for performance. Brandl et al. [40] found that very thin films in amorphous-crystalline nanocomposites can contain residual order from nearby crystals, with a transition region of finite width being observed. Amorphous complexions are slightly different from amorphous-crystalline composites, in that the former finds a local equilibrium configuration while the latter does not necessarily do so. Atomistic models of amorphous complexions have shown that SSRO motifs similar to those observed in bulk amorphous metals are present, yet the confining crystals lead to spatial gradients through the complexion thickness [41]. Garg and Rupert showed that these variations were determined by the incompatibility between the confining grains, with more ordered structural packing found at the outer edges of the complexion, which can be termed the amorphous-crystalline transition regions (ACTRs), while more disordered SSRO types are preferred in the complexion interior [42].

While both local composition and local structural order have been identified as important descriptors of amorphous complexions, their connectivity has not been investigated to date. In this study, we investigate the spatial distribution of dopant elements and SSRO within amorphous complexions in a variety of Cu-rich alloys (Cu-Zr, Cu-Zr-Nb, and Cu-Zr-Nb-Ti) using a combination of transmission electron microscopy (TEM) and atomistic simulations. By studying alloys with increasing chemical complexity, we find that different dopant species preferentially segregate to specific regions of amorphous interfacial complexions. Zr promotes amorphization and segregates with the highest concentration to the complexion interior, while Nb and Ti are found to be most concentrated at the ACTRs in ternary and quaternary alloys. To connect the local composition to SSRO, we use complementary atomistic simulations enabled by a new machine learning interatomic potential. Dopant segregation patterns in a simulated Cu-Zr-Nb alloy are in good agreement with the experimental findings, providing an important benchmark. Subsequent analysis shows that the inner region of the complexions is more disordered, demonstrating a coupling between the chemical segregation and structural short-range ordering motifs. In addition, alloy composition strongly affects the patterning of SSRO through the complexion thickness and within the plane of the ACTR, providing future alloy design guidance.

## 2. Materials and methods

### 2.1. Materials processing

Cu-3 at.% Zr (Cu-Zr), Cu-1.5 at.% Zr-1.5 at.% Nb (Cu-Zr-Nb), and Cu-2 at.% Zr-2 at.% Nb-2 at.% Ti (Cu-Zr-Nb-Ti) alloys were prepared from elemental powders of Cu (Alfa Aesar, 99%, -325 mesh), Zr (Atlantic Equipment Engineers, 99.5%, -20+60 mesh), Nb (Alfa Aesar, 99.8%, -325 mesh), and Ti (Alfa Aesar, 99.5%, -325 mesh). In subsequent text, the unit "at.%" has been removed for brevity (e.g., Cu-3Zr would be the binary alloy). The number of alloying elements was increased, moving from binary to ternary to quaternary alloy concentrations, to increase the chemical complexity at the grain boundaries as all dopants are expected to segregate [43-45]. The powders were mechanically alloyed and their grains refined using hardened steel vials and milling media in a SPEX SamplePrep 8000M high-energy ball mill inside of a glovebox under Ar with < 0.15 ppm $O_2$ levels, to limit contamination. A process control agent of 2 wt.% stearic acid (Alfa Aesar) was added to each sample to minimize cold welding. The powders were consolidated into 14 mm diameter pellets first with cold pressing at room temperature for 10 min under 25 MPa, followed by hot pressing at 950 °C for 1 h under 50 MPa. The consolidated pellets were subsequently annealed at 950 °C for 5 min and then rapidly quenched by placing the pellets onto a liquid nitrogen-cooled Al heat sink.

### 2.2. Microstructural characterization

Average sample composition was determined using an FEI Quanta 3D FEG dual beam scanning electron microscope (SEM)/focused ion beam (FIB) microscope. X-ray diffraction (XRD) measurements were obtained from a Rigaku Ultima III X-ray diffractometer with a Cu Kα radiation source operated at 40 kV and 30 mA with a one-dimensional D/teX Ultra detector and integrated software package (Rigaku PDXL). TEM specimens were prepared with the same FEI Quanta 3D SEM/FIB using a $Ga^+$ ion beam and OmniProbe, leaving the sample welded onto a Mo grid at the end. Next, the lamella was thinned on a Tescan GAIA3 SEM/FIB and then a final 700 eV polish was performed using a Fischione Model 1040 NanoMill to remove any residual surface damage. Scanning TEM (STEM) with EDS was used to identify amorphous complexions and map the local chemistry using a JEOL Grand ARM300CF with double Cs correctors operated at 300 kV and equipped with dual 100 $mm^2$ silicon drift detectors. Careful tilting of the sample was used to ensure that the complexions were in an edge-on condition. Nanobeam four dimensional scanning transmission electron microscopy (4D-STEM) was also performed on the same JEOL Grand ARM300CF operated at 300 kV, using a convergence semiangle of 1.8mrad. The electron probe was rastered over the targeted area with a step size of 1.0 nm and a dwell time of 0.04 s per step. A nanobeam electron diffraction pattern was acquired at each probe position by a Gatan OneView IS camera through Gatan STEMx.

### 2.3. Atomistic simulations

Hybrid atomistic molecular dynamics (MD)/Monte Carlo (MC) simulations within the Large-scale Atomic/Molecular Massively Parallel Simulator (LAMMPS) code [46] were used to create equilibrated amorphous grain boundary complexions. MC steps were performed in a variance-constrained semi-grand canonical ensemble [47] after every 100 MD steps while maintaining the global composition fixed at a target level. Unfortunately, no interatomic potential exists that is capable of modeling the multi-component alloys targeted in this study. As such, a new machine learning interatomic potential (ML-IAP) for the Cu-Zr-Nb system was created using the MACE package [48] based on 8692 different atomic configurations, with structural optimization and system energy calculations performed with density functional theory (DFT) using the Vienna ab initio simulation package (VASP). The projector augmented plane-wave approach [49, 50] and the Perdew-Burke-Ernzerhof exchange-correlation generalized gradient approximation (GGA) functional [51] were used. The details of the database creation and potential fitting are included in the Results and Discussion section, as these research tasks were carried out specifically for this study. This ML-IAP allowed amorphous complexion structure and chemistry to be investigated in both Cu-Zr and Cu-Zr-Nb, to probe the effect of increasing chemical complexity and chemical patterning.

A bicrystal sample was created as a representative grain boundary, with the X-axis aligned along the [110] direction of Grain 1 and the $[1\bar{1}1]$ direction of Grain 2. The two grains were oriented perpendicular to each other to provide a high degree of incompatibility. The model had approximate dimensions of 6 nm × 6 nm × 37 nm and contained 71,552 atoms. Amorphous complexions were first created in a Cu-Zr alloy using a well-known embedded-atom method (EAM) potential [30], as this potential is much more computationally efficient than our new ML-IAP. The EAM potential has been used extensively for modeling nanocrystalline and amorphous alloys in this system, including a previous study which explained how variations in grain boundary crystallography affect SSRO in amorphous grain boundary complexions [42]. The samples were first created in pure Cu and equilibrated at 1000 K using a Nose-Hoover thermo/barostat at zero pressure, and then subsequently doped with Zr using the hybrid MD/MC approach. During this simulation, grain boundary segregation and premelting occurred, resulting in a pair of amorphous grain boundary complexions. Subsequent simulations were performed with the ML-IAP to create candidate Cu-Zr and Cu-Zr-Nb alloys, allowing for dopant redistribution and local structural relaxation with the new potential. Models with grain boundary concentrations of roughly 45 at.% dopants were studied to be similar to the configurations investigated in experiments. The ratio of Nb to Zr in the complexion region was varied from 0 (Cu-45 at.% Zr at complexion) to ~0.2 (Cu-38 at.% Zr-7 at.% Nb at complexion), followed by further equilibration with MC/MD. Analysis of bicrystal models with lower complexion compositions can be found in Supplementary Note 7.

The simulations were stopped when the energy of the system converged to a stable state, signaled by the potential energy gradient lower than $10^{-4}$ eV per atom over a 20 ps period. Complete data for potential energy as a function of simulation time for the relaxation treatments using the ML-IAP are included in Supplementary Materials Note 4. Smaller metallic glass models were also used to gain insight into the complexion interior (further discussion and benchmarking is provided in the Results and Discussion section), comprised of a cubic cell with dimensions of approximately 5 nm × 5 nm × 5 nm and containing 6912 atoms. These samples were created by first melting the sample at 2000 K and then cooling to 1000 K over 10 ps. MD/MC equilibration was then performed at 1000 K following the same procedure as the grain boundary samples to allow for local structural and chemical relaxation. Grain boundary atoms were identified using common neighbor analysis (CNA) and a local entropy fingerprint was calculated [52], both using the OVITO software [53] A configurational disorder parameter was used to quantify local structure within the amorphous complexions and metallic glasses. This parameter was calculated in LAMMPS using the *orientorder/atom compute* and setting the component's value to 6 with a cutoff radius 3.0 Å [54, 55], with a value of 0 signaling a fully ordered, crystalline structure while higher values approaching and sometimes exceeding 1 signaling a disordered structure. Additional details of the disorder parameter calculation are provided in Supplementary Materials Note 5.

## 3. Results and Discussion

### 3.1. Chemical distribution within amorphous interfacial complexions

XRD scans of the mechanically alloyed, consolidated, and quenched samples (Figure S1 in Supplementary Materials) showed a face-centered cubic (FCC) primary phase and 1.5 vol.% or less content of impurity phases ZrC, $ZrO_2$, or NbC. These secondary phases were formed due to reactions with the process control agent and are typical of mechanically alloyed powders. For Cu-Zr alloys created by an identical processing procedure, we previously showed with atom probe tomography that overall O concentration was ≤0.8 at.% while overall C and N concentrations were ≤0.5 at.% [56]. The Scherrer equation was used to estimate the grain size of the primary phase from the XRD scans, with a constant grain size of ~35 nm measured. Representative bright field STEM images of the grain structures of the Cu-Zr-Nb-Ti, Cu-Zr-Nb, and Cu-Zr alloys are shown in Figures 1(a-c), respectively. TEM measurements gave average grain sizes of 40 ± 8 nm, in reasonable agreement with the XRD measurements. Grain size was calculated from at least 40 grains identified in the BF-STEM images by measuring the areas of individual grains and then converting to circular equivalent diameter. Lighter outlines around some grains are denoted by yellow arrows, with these sites being potential amorphous complexions between the grains that were further investigated with high resolution imaging. Figure 1(d) is a high resolution bright field STEM image of an amorphous grain boundary complexion in the Cu-Zr alloy. Lattice fringes are clearly observed on each side of the boundary, outside the dashed white lines that denote the outer edges of the complexion. Data from an EDS line scan of the same grain boundary is overlaid on the image, where it is observed that Zr segregation is highest in the interior region of the complexion. A gradient is observed over ~1-2 nm near the ACTR, where the complexion meets the confining crystals. The observation of Zr depletion at the ACTR provides experimental confirmation of earlier predictions from atomistic simulations for this same binary alloy [42].

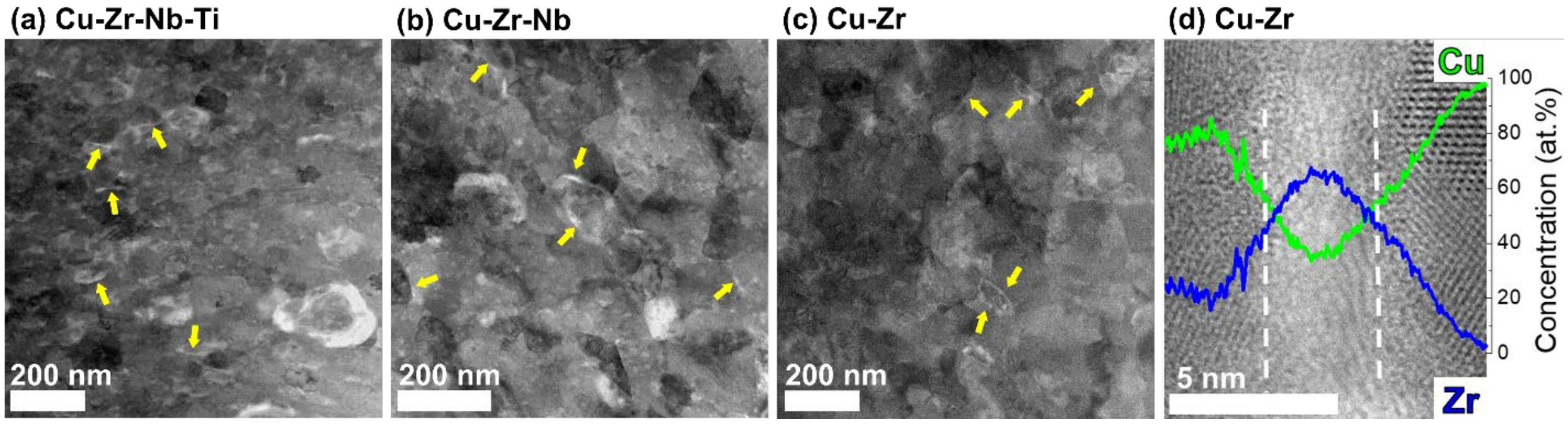

**Figure 1. Bright field STEM images of nanocrystalline grain structures in the (a) Cu-Zr-Nb-Ti, (b) Cu-Zr-Nb, and (c) Cu-Zr alloys. Yellow arrows denote boundaries enriched with heavier dopant elements, which were targets for further investigation. (d) A higher-magnification bright field STEM image of an amorphous complexion in Cu-Zr with corresponding EDS line scan superimposed shows Cu-depletion (green line) and non-uniform Zr-enrichment (blue line).**

In Figure 2, two representative examples of amorphous complexions at grain boundaries in Cu-Zr-Nb and Cu-Zr-Nb-Ti are shown. The bright field STEM images in Figure 2(a) demonstrate that the complexions are structurally disordered. The EDS maps in Figure 2(b) show that the dopants segregate to the amorphous complexions while Cu is significantly depleted. For the Cu-Zr-Nb example, it is clear from the line scans in Figure 2(c) that Cu is depleted and Zr is enriched with the highest concentration in the middle of the complexion. To observe the Nb segregation behavior more clearly in this sample, this data is presented by itself in Figure 2(d). Nb generally segregates to the complexion, yet a local minima is observed in the middle of the complexion. Notably, the highest Nb concentrations appear along the outer edges, at the ACTRs. Similar behavior is also seen in the Cu-Zr-Nb-Ti sample presented in the bottom row of Figure 2, where Zr is again observed to segregate most strongly to the middle of the complexion while Nb and Ti both have local maxima at the ACTRs. While the focus of this study is on planned metallic dopants, a qualitative picture of impurity element distribution can also be obtained from EDS scans. Figures S2-S4 show spatial distributions of both metallic and impurity elements across representative complexions in the three alloys, with evidence of O and C enrichment of the grain boundary observed.

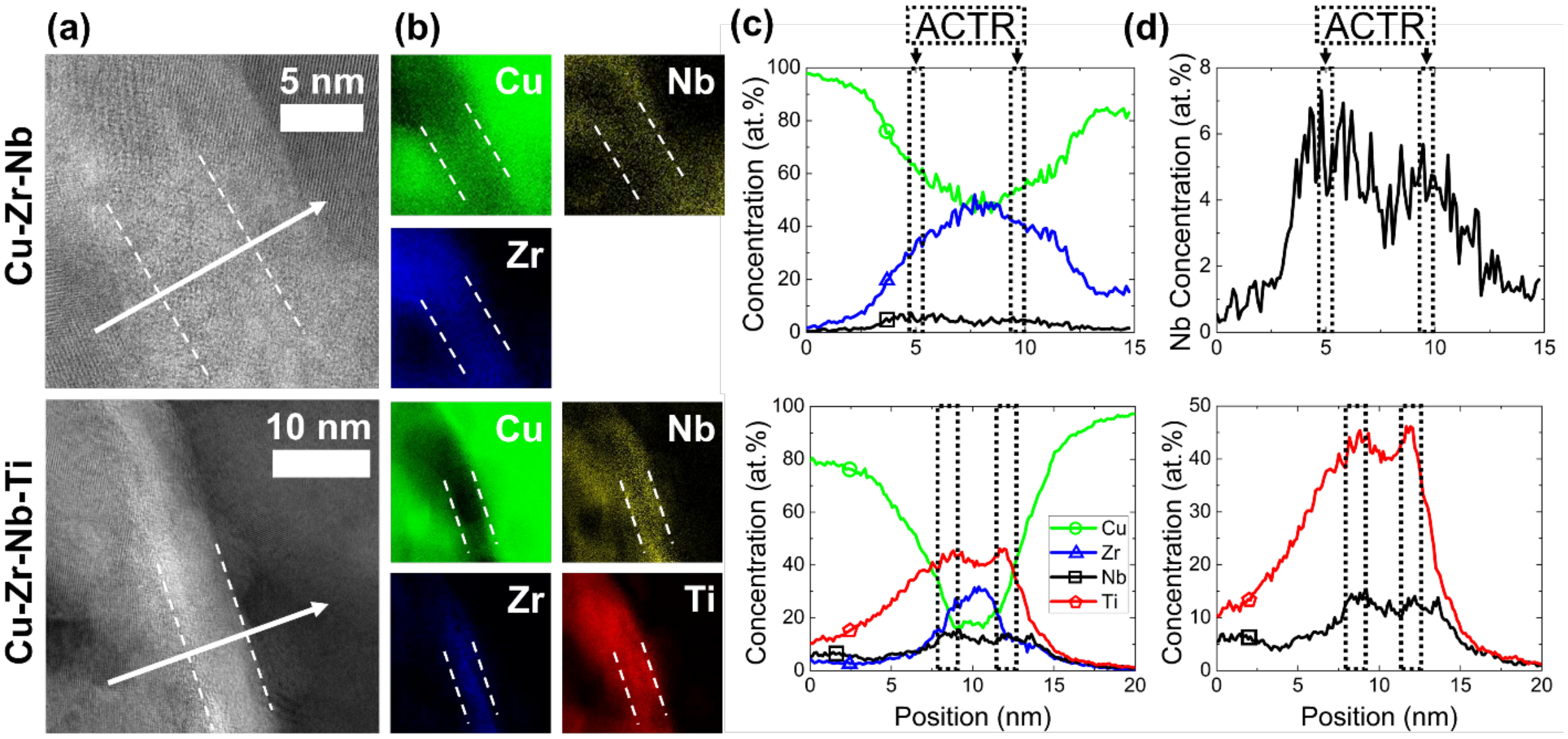

**Figure 2. (a) Bright field STEM images and (b) corresponding EDS maps of amorphous complexions in Cu-Zr-Nb and Cu-Zr-Nb-Ti. (c) EDS line scans were obtained along the white arrow in (a), where the black dashed boxes indicate the amorphous-crystalline transition regions (ACTRs). (d) Magnified views of the Nb and Ti line scans.**

While our primary analysis focuses on amorphous complexions between Cu grains, amorphous intergranular films were also observed at other types of interfaces such as interphase boundaries between smaller (~10-20 nm) secondary phase particles and Cu-rich grains. In Figure 3(a), the circular Cu-depleted region corresponds to a dopant-rich (Zr, Nb, and Ti) particle, labeled *P*, with an amorphous complexion (outlined by dashed white lines) found at the interface separating the particle and a Cu-rich grain, labeled *G*. This complexion is primarily Nb with some small amounts of the Zr and Ti content segregated to the lower left of the complexion. In Figure 3(b), an amorphous complexion is found between another dopant enriched, Cu-depleted particle and a Cu-rich grain. This complexion has a relatively uniform Ti concentration, with Nb and Zr enrichment at the edge nearest the particle and Cu enrichment at the edge of the complexion touching the grain. Unfortunately, the high curvature of the interfaces shown in Figure 3 makes it impossible to achieve the edge-on conditions necessary for rigorous quantitative analysis. While not quantifiable from EDS, C and O segregation were observed in the EDS maps overlapping with the dopants. Amorphous interphase complexions have been observed in other materials, such as between Cu and Sn layers which were sputtered to create an initial interface free of intermetallic compounds and then subsequently exposed to electromigration [57].

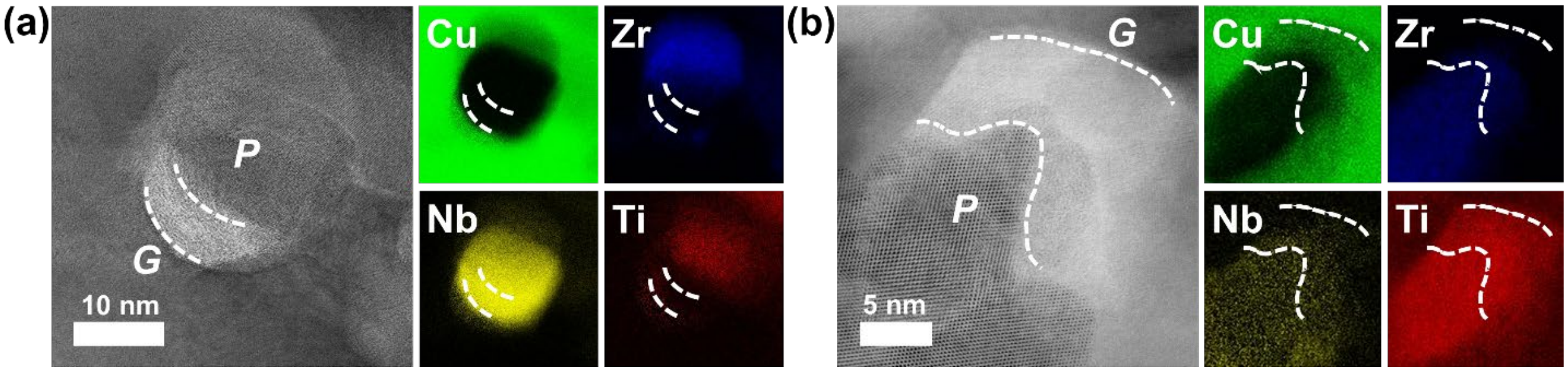

**Figure 3. Bright field STEM images and corresponding EDS maps of two different Cu-depleted and dopant-enriched particles in the Cu-Zr-Nb-Ti sample. Both exhibited amorphous interphase complexions along an interface with the Cu-rich matrix phase, although the dopant segregation behaviors are different.**

For quantitative analysis of the amorphous complexions between two Cu grains, we limited our investigation to edge-on grain boundaries with minimal curvature and significant enrichment of the dopant species, subjectively defined here as having at least 50 at.% total dopant concentration. Although a few complexions initially appeared to have lower dopant concentrations, there was always evidence of overlapping grains (found by over or underfocusing) or secondary phases in those cases which obscured the measurements. Two distinct regions were defined for quantification that could potentially have different atomic packing: (1) the ACTR and (2) the complexion interior. The ACTR starts where the lattice fringes meet the edge of the complexion and continues for a short distance into the complexion. A region with a constant width of 0.4 nm, chosen to capture several data points from the EDS line scans and to be similar to the ACTR thicknesses predicted by prior atomistic simulations [41, 42], is centered on the ACTR to determine the average of the concentration in that region. The ACTRs are labeled on Figures 2(c) and (d) with black dashed boxes. A region with the same width is defined at the center of the interior region to allow for the capture of a consistent EDS signal. By defining a constant width of 0.4 nm for the analysis region of the ACTR and the interior, a very similar volume of the sample is measured, which enables a clean comparison of the concentrations. Differences in TEM sample thickness can be a source of potential error in EDS measurements. Here, we restrict any comparisons of composition to different regions within individual complexions, meaning the measurements are taken at most a few nanometers apart to minimize this issue. Light elements should also only be discussed in qualitative terms, so we restrict any quotations of local concentration to the heavier metallic elements in the alloys. The ACTR and complexion interior regions are denoted with blue dotted boxes on the bright field STEM image in Figure 4(a), with the red dashed lines indicating where the lattice fringes end.

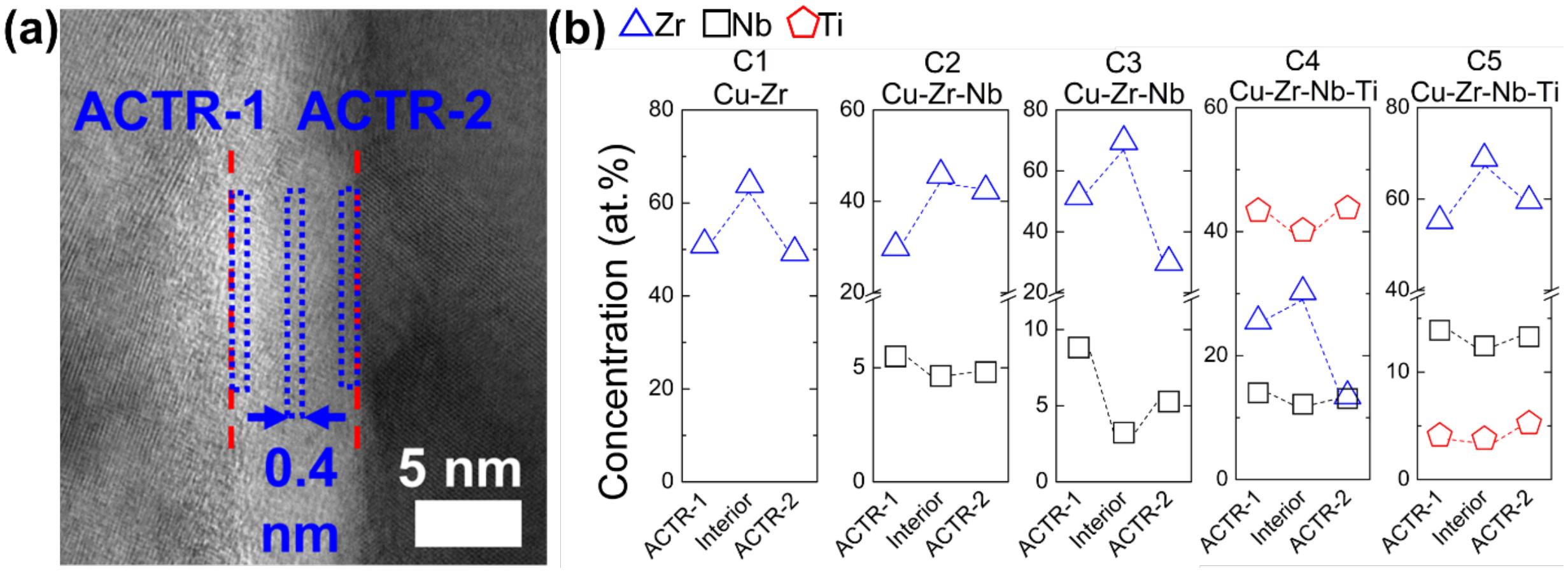

**Figure 4. (a) Bright field STEM image of an amorphous grain boundary complexion in Cu-Zr-Nb-Ti, where blue dotted boxes denote the ACTR and complexion interior regions targeted for EDS measurements and the**

**red dashed lines show where the lattice fringes end. (b) Average concentrations for different regions within amorphous grain boundary complexions from (C1) Cu-Zr, (C2, C3) Cu-Zr-Nb, and (C4, C5) Cu-Zr-Nb-Ti.**

The averages of the data points contained within the ACTR and interior regions are plotted in Figure 4(b) for each of the complexions. One complexion is shown for Cu-Zr, two for Cu-Zr-Nb, and two for Cu-Zr-Nb-Ti, with the abbreviations C1-C5 used to mark each instance. In all of the complexions, Zr concentration is highest at the center of the amorphous complexion. The average Zr content of the ACTRs is 41 at.% whereas the complexion interiors have an average of 56 at.% Zr (we note that these numbers are heavily influenced by C4, which is heavily doped with Ti). In contrast, Nb and Ti concentrations are higher at all eight ACTRs in the multi-component alloys as compared to their respective complexion interiors. Sometimes the enrichment is modest, such as for Nb in C2 and C4. However, large levels of enrichment are observed for Nb in C3 and Ti in C4. While the exact level of segregation of each dopant varies between the different complexions, the relative trends of enrichment or depletion in ACTR and complexion interior regions are consistent across all samples. ACTRs are enriched in Nb and/or Ti, while the complexion interiors have higher relative Zr concentrations. The uncertainty in elemental composition associated with each of the EDS maps was ≤0.16 at.% for each element, as reported in the Gatan DigitalMicrograph software. This aligns with the typical relative error of ±2-5% of the quoted values for standardless quantitative EDS analysis [8].

The segregation trends described above can be understood in the context of structural packing variations within the amorphous complexions. The ACTRs are generally more structurally ordered than the complexion interior, due to the confining grains nearby. Thus, Nb and Ti appear to prefer the more ordered ACTRs, while Zr prefers the disordered complexion interior. Prior reports have shown that when added to Cu by themselves, the three dopant species have different segregation behaviors, which can provide insight into their preferred site types. Zr promotes amorphous complexion formation, with the earliest examples of amorphous complexions in nanocrystalline alloys found in Cu-Zr [60]. In contrast, Nb and Ti tend to form ordered segregation patterns at grain boundaries. Nb is nearly insoluble in solid Cu [61, 62], resulting in strong Nb segregation and the stabilization of ordered grain boundary types [63-65]. Specifically, Nb atoms segregate to grain boundary sites, fill into the grain boundary as dopant concentration is increased, and eventually nucleate precipitates. With high enough Nb concentrations, the heterogeneous nucleation process leads to the precipitates transforming to ordered Nb films [63]. Similarly, Ti doping of Cu results in ordered grain boundary segregation in diffusion barrier layers [66] and sputtered thin film alloys [67]. Ti solute atoms diffuse rapidly to grain boundaries upon heating and, analogous to Nb, form an intermediate ordered secondary phase. With increased annealing, the Ti secondary phase grows into an interfacial layer. Given that Nb and Ti prefer ordered grain boundary sites, the partitioning observed in Figure 4(b) suggests that the more ordered ACTR draws these elements in. Elements preferring to segregate to the edges of grain boundaries has been observed for other systems, such as Bi-enrichment at the edges of ZnO crystals, with a clear correlation showing that the edges have disorder parameter values between the purely ordered crystals and purely disordered grain boundary interior [68].

### 3.2. Development and benchmarking of the ML-IAP

Armed with an understanding of where the various dopants segregate, it is important to ask how the increased chemical complexity of the multicomponent amorphous complexions leads to changes in the SSRO distribution. The most direct way to investigate this question would be to directly measure the SSRO in the different regions of the complexions, with nanobeam electron diffraction providing a possible route for doing so. This method has been used in the literature to probe the SSRO of bulk metallic glasses. For example, Hirata et al. found evidence of widespread frustration of the icosahedral packing in a Zr-Pt metallic glass using nanobeam electron diffraction [69]. To provide experimental confirmation of structural differences within the complexion, nanobeam electron diffraction patterns were collected across an amorphous complexion in Cu-Zr and are presented in Figure 5. Diffraction patterns taken from ACTR-1, the interior of the complexion, and ACTR-2 are shown in Figures 5(a)-(c), respectively. A bandpass mask was applied in the Gatan DigitalMicrograph software to block the outer diffraction spots, such that only the halos around the central beam spots were visible. The gamma value was increased to more clearly show the contrast of each halo. The broad and diffuse halos observed around the central beam spots indicate the disordered structure of the three regions. Similar observation of diffraction halos was made by Zhu et al. for a Zr-Cu-Al metallic glass, where the halos maintained a circular symmetry [4]. Difference images between ACTR-1 and ACTR-2 with respect to the complexion interior are shown in Figures 5(d) and (f), respectively. ACTR-1 has specific locations where significant differences are found, with four examples denoted with green arrows. In contrast, ACTR-2 has differences from the interior yet they are more diffuse across the diffraction pattern around the main halo. Importantly, these images show that (1) the ACTRs are different from the complexion interior and (2) the two ACTRs are different from one another. Additional nanobeam diffraction patterns are shown in Supplementary Figure S5 with discussion on analysis complications for these samples. Nanobeam electron diffraction has been used previously to demonstrate the structure of amorphous materials including metallic glasses [5, 6]. For example, Francis and Voyles recently developed a method to filter patterns, identify 4-D diffraction vectors, and cluster the vectors into characteristic diffraction patterns to separate different structural motifs from an amorphous thin film sample [7]. While Figure 5 provides qualitative evidence of structural differences, complementary simulations were pursued to more quantitatively study of the effect of multi-component doping on SSRO within amorphous complexions.

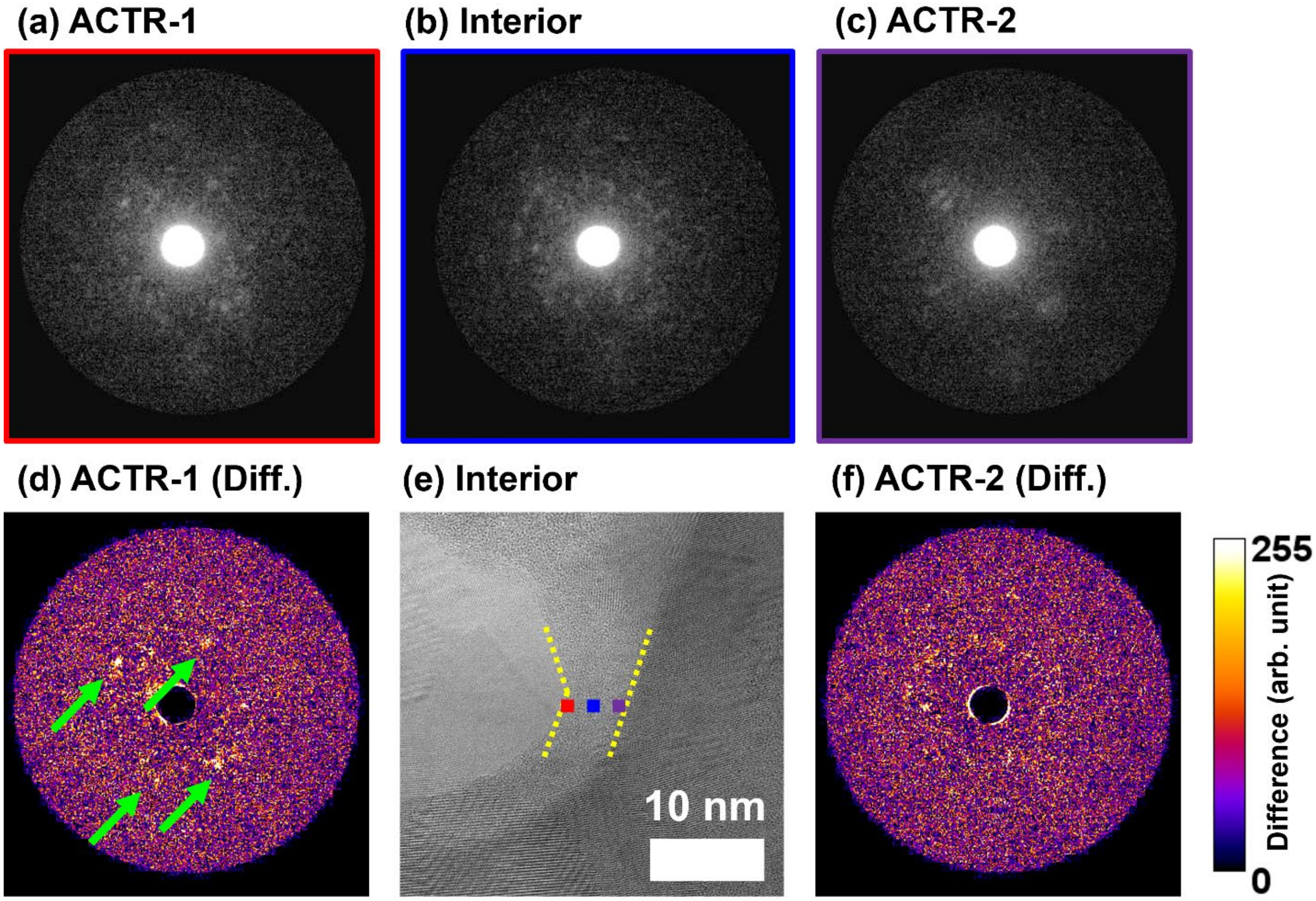

**Figure 5. Nanobeam electron diffraction patterns taken from (a) ACTR-1, (b) the complexion interior, and (c) ACTR-2 in a complexion in Cu-Zr. The diffraction patterns all have diffuse halos around the central beam spot, demonstrating structural disorder. (e) The amorphous complexion is outlined by yellow dotted lines, with colored squares showing where each diffraction pattern was taken. Difference images for (d) ACTR-1 and (f) ACTR-2, taken with respect to the complexion interior, show that there are structural variations within the complexion.**

Atomistic simulations offer an alternative method of investigating amorphous complexions where local chemistry and structure can be easily quantified. As no existing interatomic potential could treat the alloys of interest, a new ML-IAP was created for this task. Database selection is critical for the development of a meaningful ML-IAP that can simulate the correct atomic bonding between elements in both bulk phases and defect configurations. Here, heavily doped and defected structures are of primary importance. First, the stable phases of the Cu-Zr-Nb system were added to the database. All phases for this system available in the Materials Project [70] were used to establish the lower limit on formation energy for the system (Figure S9). The energies of these phases were calculated at 0 K with DFT. To include phases relevant to finite temperature conditions, equilibrium phases predicted by Thermo-Calc were also included [71]. Examples of phase diagrams used to extract phases and their energies are shown in Figure S10-S12. In addition to ordered phases, solid solutions with FCC, body-centered cubic (BCC), and hexagonal close packed (HCP) structure were included to ensure that a range of local chemical environments were treated. Additional calculations were performed on crystalline and amorphous specimens at elevated temperature of 800-5000 K using ab initio MD in VASP. In total, 8692 atomistic configurations were used to create the fitting database for the ML-IAP, with examples of crystalline and defected states shown in Figure 6(a). Structures derived from phase diagrams, doped Cu alloys, and solid solutions are all captured in the database and function as a priori knowledge for amorphous structure formation. Figures 6(b) and (c) present the formation energy distribution and energy histogram, respectively. Additional description of the ML-IAP database creation can be found in Supplementary Materials Note 6. Figure S13 shows the results on training and validation datasets for the ML-IAP, where good performance is observed with small root mean square errors of 5.4 and 5.9 meV/atom for energy. For forces, the component-wise RMSE values are 86.8, 86.2, and 85.2 meV/Å for the x-, y-, and z-directions in the training set and 86.2, 85.6, and 83.7 meV/Å for the corresponding directions in the validation set.

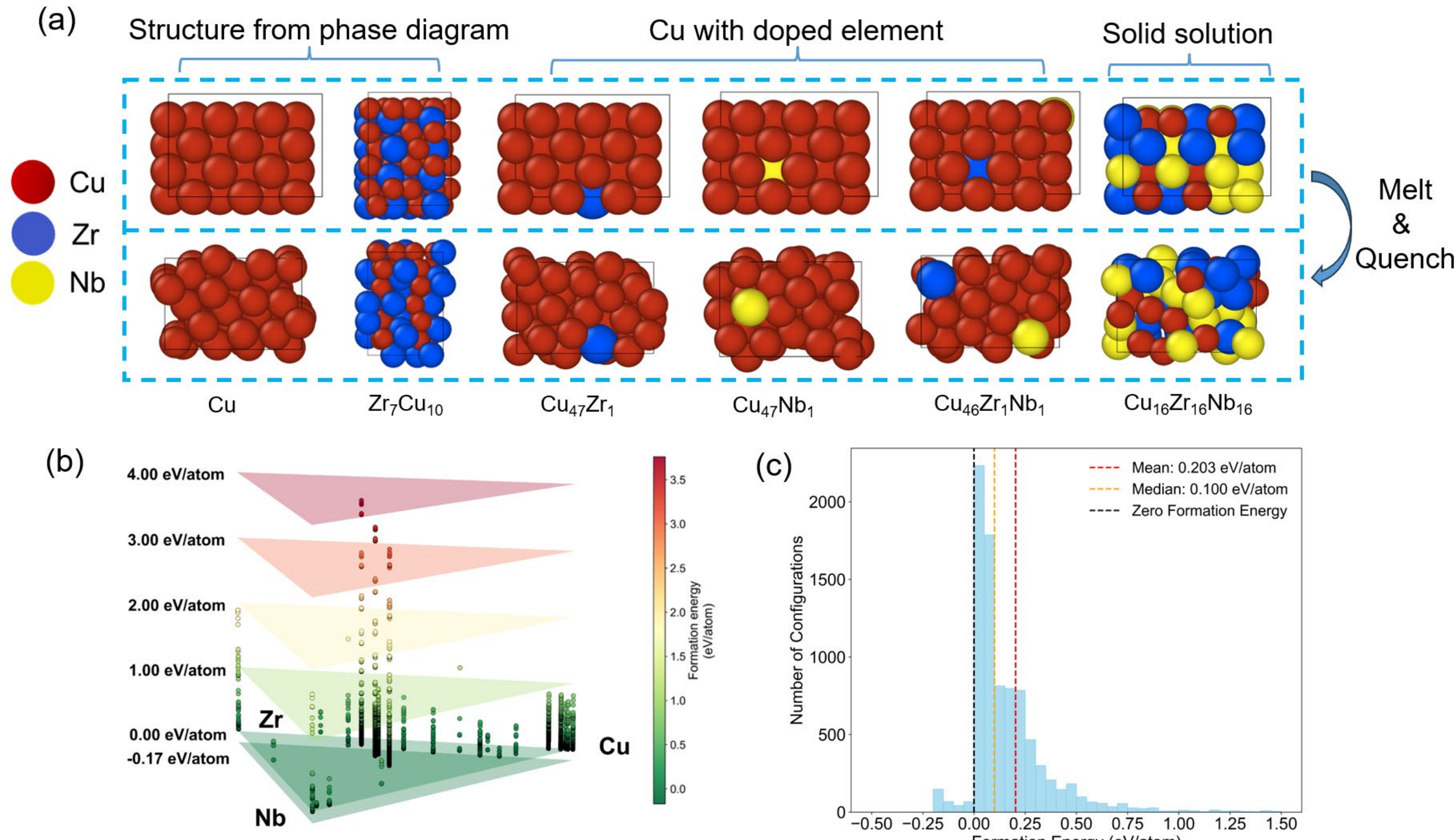

**Figure 6. (a) Examples of crystalline and defected atomic configurations used in the fitting database for the ML-IAP. (b) Graphical distribution and (c) histogram of formation energies for the atomic configurations.**

In addition to general validation of the ML-IAP against DFT calculations, our STEM observations of grain boundary segregation and chemical patterning allow the potential to be benchmarked against the exact behavior that is being investigated. Figure 7(a) shows an illustration of simulated Cu-Zr-Nb bicrystal sample that has been equilibrated at 1000 K with MD/MC simulations. Both dopant species are clearly observed to segregate to the grain boundaries, matching the first behavior of interest. Figure 7(b) shows the overall complexion structure, with the ACTRs and complexion interior labeled. The regions that are identified as non-crystalline with CNA define the grain boundary (Figure 7(c)), with 10% of the GB thickness (0.5 nm) regions near the edges defined as the ACTRs (Figure 7(d)) to be consistent with our experimental measurements shown in Figure 4. Defect mesh analysis within OVITO provided the outer edge of the interfacial region, with the ACTRs beginning 0.264 nm inward (corresponding to the nearest Cu-Cu bond distance at 1000 K) to ensure that no FCC grain interior atoms were included. Another 10% of the GB thickness (0.5 nm) region in the center of the complexion is defined as the complexion interior for any quantification, again to match our experimental measurements. To further match the composition from experiment, more Zr was randomly assigned as described in the Methods section. By using same definition of grain boundary region, Figure 8 presents concentration as a function of position, moving across one amorphous complexion in each of the modified Cu-Zr and Cu-Zr-Nb alloys. All of the complexions are depleted of Cu and enriched with the dopant species, as shown in the top row of Figure 8. The bottom row of Figure 8 provides a detailed view of the dopant concentrations across the complexion thickness, where the ACTRs are labeled in grey (10% of the GB thickness, 0.5 nm). For all of the alloys, the Zr concentration is always higher in the complexion interior than in the ACTRs. Looking specifically at the multi-component Cu-Zr-Nb alloys, the Nb concentration always shows a peak value within the ACTRs, with slightly lower but still enriched levels in the complexion interior. The spatial patterning of dopant concentration therefore matches the experimental observations in Figures 1, 2, and 4, demonstrating that the ML-IAP is able to capture the grain boundary behavior of interest. It is important to note that this outcome was not predetermined, as none of the experimental observations or even direct simulations of boundaries were used for creation of the ML-IAP. Rather, a range of crystalline and disordered states was used to ensure that structural state space was randomly sampled. This procedure indeed captures the interaction of local structure and chemistry in a way that the complexion formation and subsequent chemical patterning within the complexion is a natural consequence.

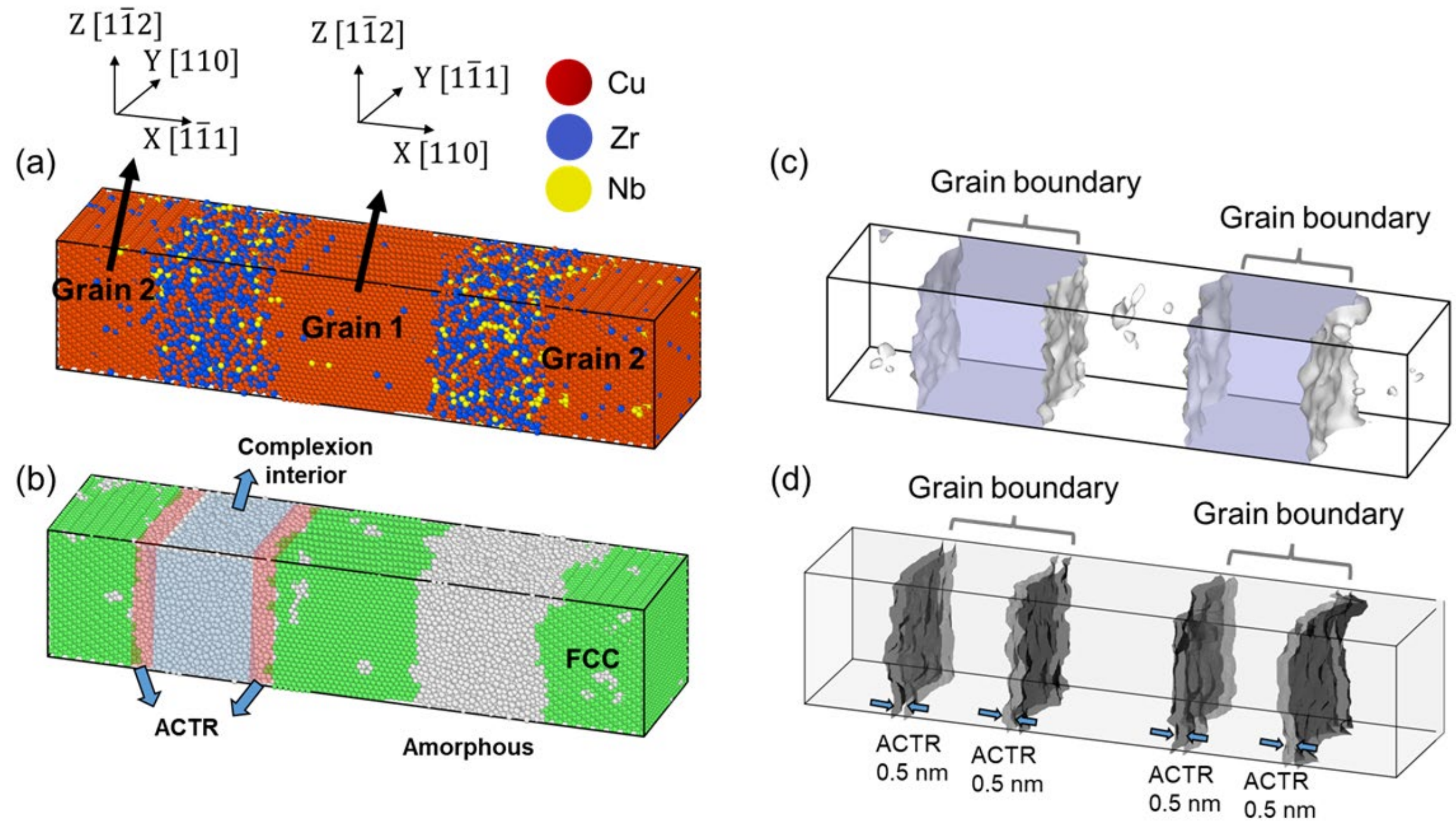

**Figure 7. (a) A representative example of a Cu-Zr-Nb sample with two grain boundaries, where the two grains are oriented perpendicular to one another to give a high degree of incompatibility. (b) Complexion interior and ACTR (~10% thickness of the complexion) locations are defined within the complexion (c,d) in a manner that matches the experimental technique.**

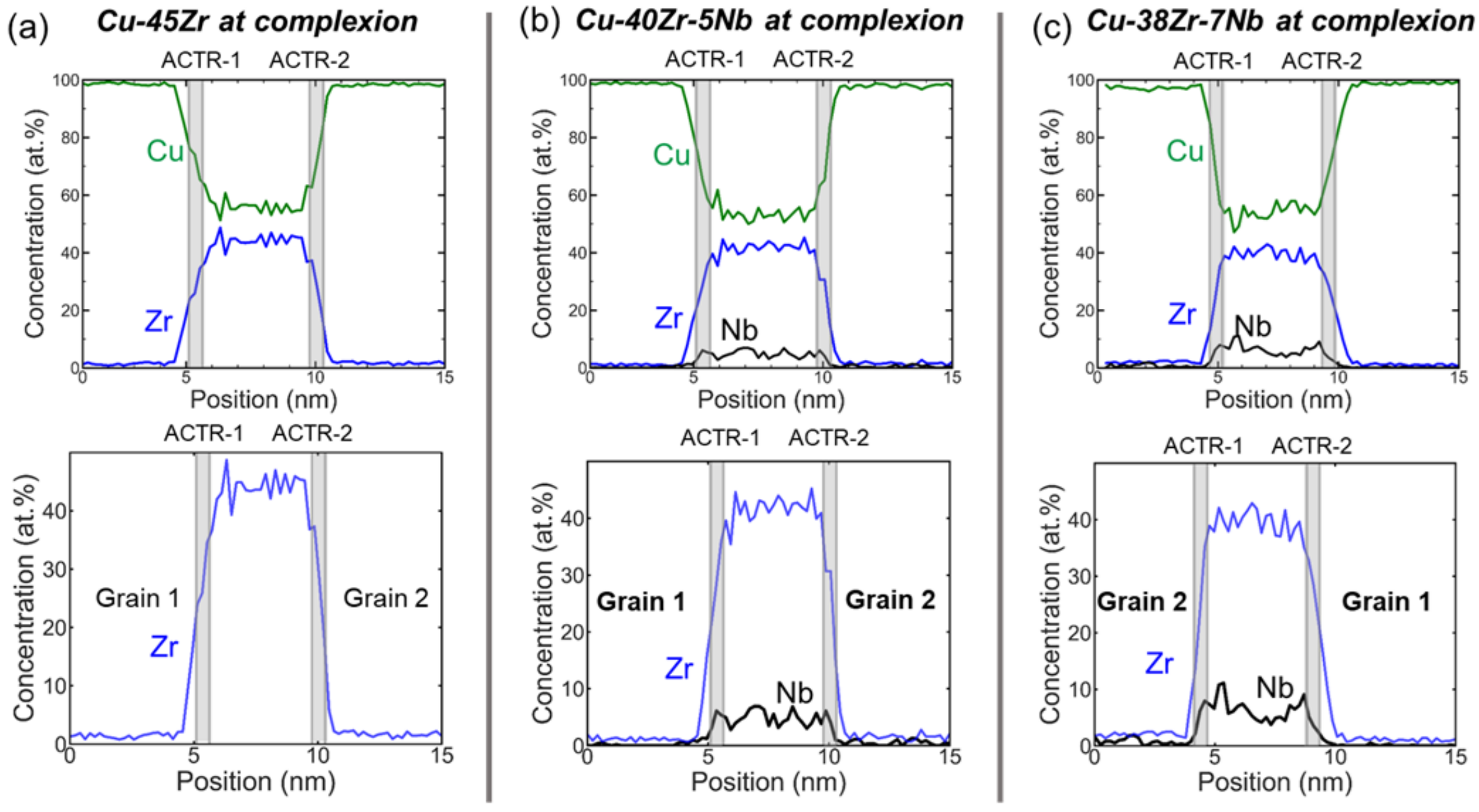

**Figure 8. Chemical concentration maps across Zr content modified amorphous grain boundary complexion with complexion composition of (a) Cu-45Zr, (b) Cu-40Zr-5Nb, and (c) Cu-38Zr-7Nb. The dopants are enriched in the complexion region for all alloys. Zr concentration is highest in the complexion interior while Nb segregate in the ACTRs (denoted by grey shading), matching the experimental observations of this study.**

### 3.3. Structural variations within multi-component amorphous complexions

The accurate simulation of the amorphous complexions and their chemical patterning means that important features are being predicted, opening the door for the investigation of SSRO. Figure 9 presents measurements of the local disorder parameter for each of the four alloys studied here, with measurements taken in both the ACTRs and complexion interior. In all cases, the complexion interior is more disordered (i.e., has a higher disorder parameter) than the ACTRs, with this effect being strongest in the alloy whose complexion composition is Cu-40Zr-5Nb (Figure 9(b)). Prior work has shown that the increased order at the ACTRs is caused by the confining crystals [41] and the degree of order can be predicted by the incompatibility between the grains [42], although these earlier studies used the density of specific Voronoi packing motifs as a signal of SSRO rather than quantifying the disorder parameter. An asymmetry can be observed in Figure 9 between ACTR-1 and ACTR-2 for all samples, with the largest difference again for the specimen with a complexion composition of Cu-40Zr-5Nb, where ACTR-2 has a significantly higher disorder parameter than ACTR-1, although again the magnitude of the asymmetry is material-dependent. The asymmetry is smaller for complexions with content of Cu-45Zr and Cu-38Zr-7Nb yet still present. Garg and Rupert [14] showed that such asymmetry in ACTR structure can result in different mechanical damage tolerance when dislocations are impinging on the complexion from the different directions. Interpreting our results through this lens, one would predict that this amorphous complexion orientation would be more damage tolerant as dislocations are impinging on ACTR-1 in Figure 9(b). In general, adding more Nb increases the disorder parameter of the complexion interior as well. Generally, Figure 9 highlights spatial heterogeneity in SSRO in a direction normal to the grain boundary plane which should affect how plasticity is shared between the various regions within the amorphous grain boundary complexion.

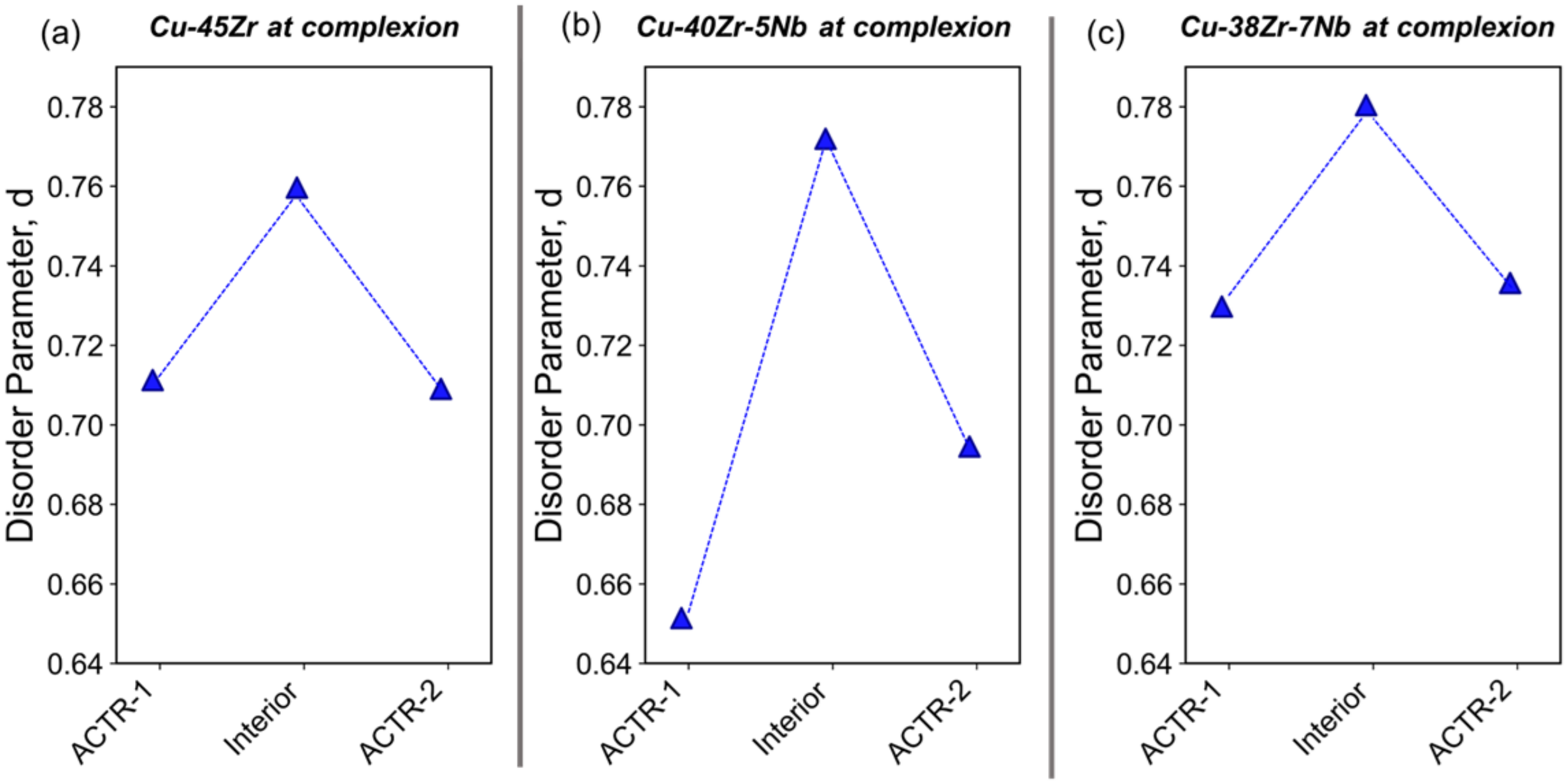

**Figure 9. Disorder parameters for different locations within an amorphous grain boundary complexion with complexion composition of (a) Cu-45Zr, (b) Cu-40Zr-5Nb, and (c) Cu-38Zr-7Nb. The complexion interiors were more disordered than the ACTRs, and the level of asymmetry in disorder parameter for the ACTRs was dependent on the alloy concentration.**

As disorder parameter is calculated for each atom, this metric of local SSRO can be further examined in terms of each element, as shown in Figure 10. A number of overall trends can be found which extend across all alloys. First, the Cu atoms in the ACTRs tend to be located in more ordered environments, yet be more disordered in the complexion interior. The relative difference appears to be sensitive to the alloy system, ratio of dopant species, and the exact region of interest. For example, the elemental variations are generally largest in the sample whose complexion composition in Cu-40Zr-5Nb. ACTR-1 shows a much broader range of values for Cu-40Zr-5Nb (Figure 10(b)) as compared to the same region for Cu-38Zr-7Nb (Figure 10(c)). Figure 11 provides a closer look of ACTR-1 in these two samples, presenting views of local disorder parameter within the ACTR plane. When comparing the complexions with content of Cu-40Zr-5Nb (Figure 11(a)) to the Cu-38Zr-7Nb (Figure 11(b)), one can see that the SSRO in the former is significantly more heterogeneous. Figure 11(c) present histograms of this same data, where the complexion with Cu-40Zr-5Nb composition is found to have more sites with low disorder parameter (i.e., relatively ordered sites). Such structural heterogeneity can actually be a positive feature and has been shown to enable more stable plasticity in metallic glasses. For example, Ma and Ding [38] showed that structural heterogeneity can cause the formation of many small shear bands, which distribute the plastic flow into many regions rather than allowing it to localize into one. Recent work from Wu et al. [73] demonstrated how planned doping with nonmetallic elements such as O, B, or C can be used to accomplish this type of structural heterogeneity and improve plasticity. Notably, these authors found that this increase in plasticity came with a simultaneous increase in strength, avoiding a property trade-off. Combined with our earlier observations, it is clear that multi-component doping can induce structural heterogeneities both normal to and within the grain boundary plane. In other words, the amorphous complexions themselves contain an internal microstructure, with local chemistry representing a powerful pathway for manipulating SSRO.

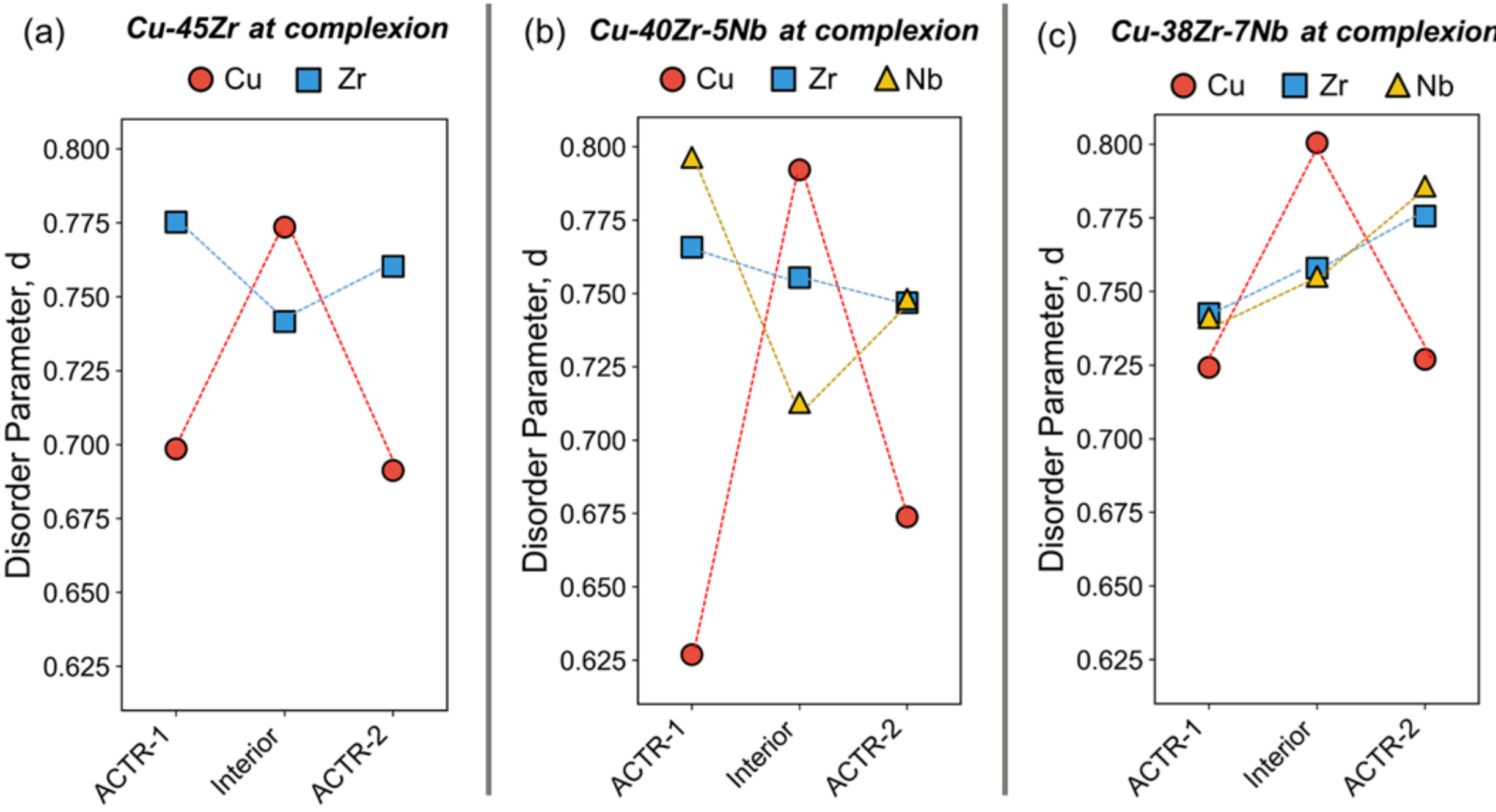

**Figure 10. Elemental disorder parameters for different locations within an amorphous grain boundary complexion with complexion composition of (a) Cu-45Zr, (b) Cu-40Zr-5Nb, and (c) Cu-38Zr-7Nb. Cu atoms generally reside in more ordered environments than the dopant species, with the relative variations being alloy dependent.**

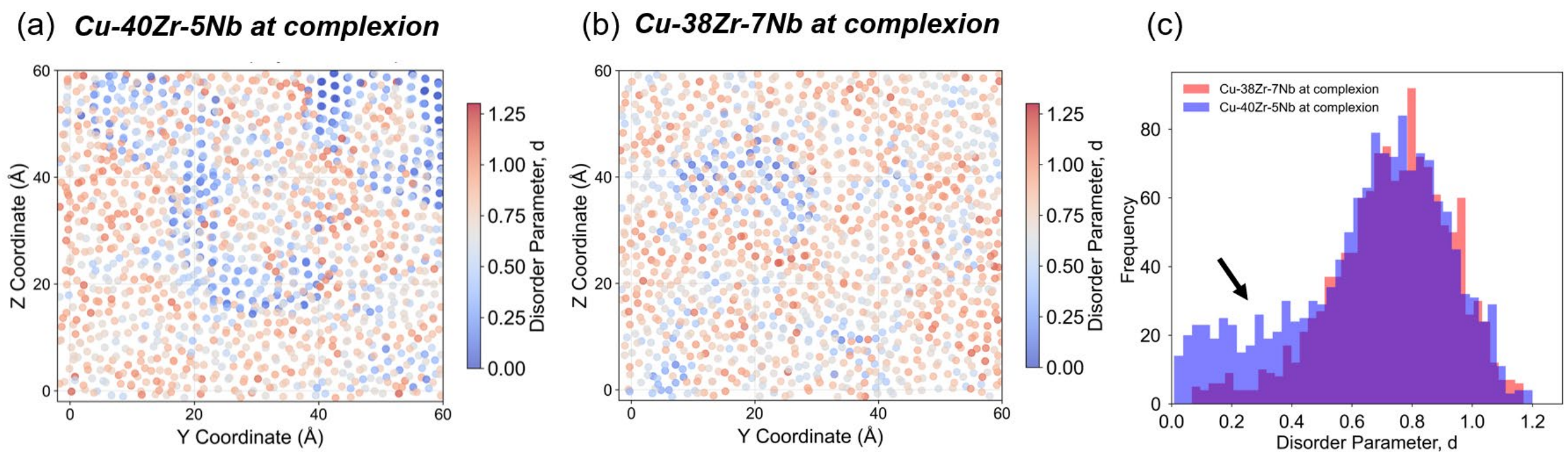

**Figure 11. In-plane distribution of the atomic disorder parameter for ACTR-1 with complexion composition of (a) Cu-40Zr-5Nb, and (b) Cu-38Zr-7Nb. Histograms of the disorder parameter distribution for the two alloys, demonstrating that a collection of relatively ordered sites are present in the complexion with Cu-40Zr-5Nb composition.**

Focusing on the complexion interior, which will also be involved in the damage process for amorphous complexions, Figure 9 shows that the ternary alloys generally have higher disorder than the binary alloy. The level of the disorder in the complexion interior is also seen to be dependent on the ratio of the dopant species, motivating further study of the effect of dopant ratio and overall dopant concentration. Bicrystal models are relatively inefficient for exploring composition space, as each specimen requires long equilibration simulations to allow for segregation and complexion formation. Smaller metallic glass models with different chemical compositions were used here to more clearly demonstrate how composition and SSRO are related. These models are meant to be most directly comparable to the complexion interiors, which are far away from the confining crystals and therefore hypothesized to not be affected. A bulk metallic glass model with a composition of Cu-45Zr-5Nb was created by melting and quenching, to be compared to the interior region of the amorphous complexion in the alloy shown in Figure 9(b) whose grain boundary composition was Cu-40Zr-5Nb. The disorder parameters for these two regions are very similar at 0.747 and 0.755 (with the small deviation likely the result of a modest composition difference), confirming that the bulk metallic glass models are reasonable analogs for the interior regions of the amorphous complexions. Following this validation, a battery of bulk metallic glass specimens were created and equilibrated with different overall dopant concentration and dopant ratios (Nb/Zr). The alloy compositions shown in Figure 12 (overall dopant concentration of 40-60 at.%) were chosen to be in the general range of the grain boundary concentrations measured experimentally here. The datasets shift downwards as dopant concentration is increased. Within a given overall dopant amount, increasing the ratio of Nb to Zr leads to increased disorder. We note that the concentrations probed here are on the Zr-rich side of the spectrum, going from only Zr as a dopant (ratio of 0) to an equiatomic doping level (ratio of 1). Prior work has shown that local structural state can impact the plasticity and damage of glassy materials. For example, Şopu et al. [74] showed that a lower degree of SSRO facilitates the brittle-to-ductile transition in metallic glass nanowires. Similarly, Yang et al. [75] demonstrated this relationship quantitatively in $Ni_{80}P_{20}$ metallic glasses, showing that samples with more disorder (achieved through structural rejuvenation) exhibit enhanced fracture energy of 1047 J/cm³, compared to 777 J/cm³ in more ordered specimens. With these prior findings in mind, Figure 12 suggests that a relatively low overall dopant concentration with a dopant ratio of 1 may offer higher disorder and damage tolerance for the amorphous complexion interior.

**Table 1. Comparison of local composition and disorder metrics for the complexion interior and a bulk metallic glass sample.**

| Sample | Local Composition (at.%) | Disorder Parameter (unitless) |
| --- | --- | --- |
| Complexion Interior | Cu-40Zr-5Nb | 0.747 |
| Bulk Metallic Glass | Cu-45Zr-5Nb | 0.755 |

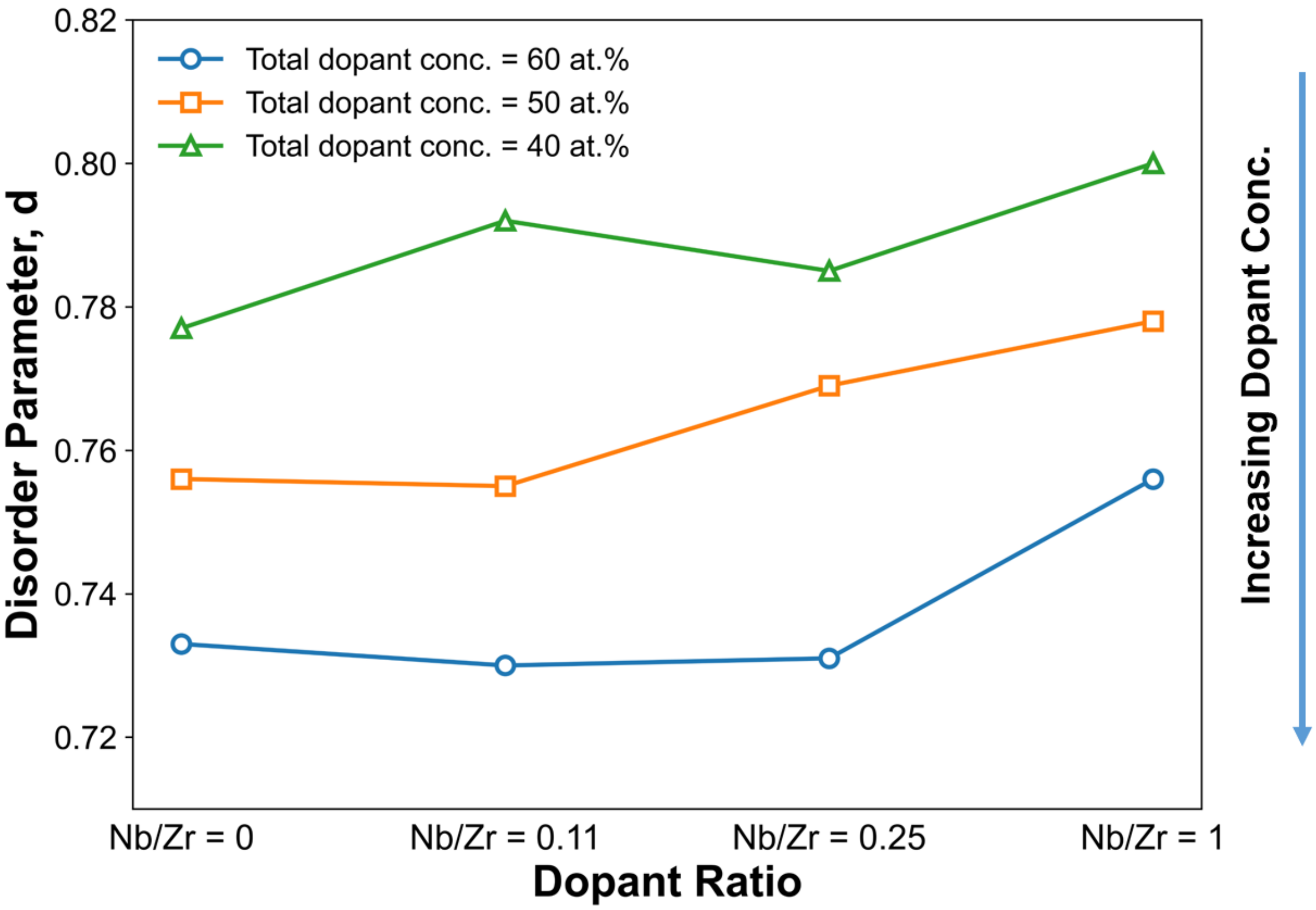

**Figure 12. Disorder parameter for bulk metallic glass specimens (comparable to amorphous complexion interiors) for different overall dopant concentrations and dopant ratios. Lower overall dopant concentrations and higher Nb/Zr ratios give higher disorder parameter values.**

Prior research has shown that amorphous grain boundary complexions can improve mechanical properties. For example, amorphous boundaries in nanocrystalline Co-Al enabled a nearly 4× increase in yield strength compared to the bicrystal and bulk single crystal counterparts [15]. Similarly, amorphous grain boundaries in nanocrystalline Ni blocked dislocations, leading to a ~33% increase in strength [16]. Here, we study amorphous complexion chemistry and structure in more detail, finding an obvious connection between SSRO and local segregation, which opens the door for the manipulation of amorphous complexion response. The ability of an amorphous complexion to absorb dislocations without cracking has been shown to be very sensitive to SSRO variations [14]. For example, Phan et al. [17] investigated the plasticity of amorphous-crystalline composites, finding that SSRO was important for determining the location of shear activation zones induced by dislocation absorption. Controlling SSRO by adding additional dopant elements can thus further impact the dislocation absorption and increase strength. Ternary and higher complexity metallic glasses are generally reported to be stronger and tougher than binary counterparts [78, 79]. For example, the addition of Al or Ag was found to increase the glass-forming ability, strength, and ductility of Cu-Zr glasses [78, 80]. Additional doping can lead to atomic-scale heterogeneities, such as Zr-rich clusters centered by Ag atom pairs in Cu-Zr-Ag [81], thus increasing the opportunity for efficient atomic packing and energy minimization. Based on the results of this study, we propose that the increased chemical complexity and spatial variations in SSRO in multi-component amorphous grain boundary complexions can be used to break-up the ordered structural types that persist due to the crystals and which drive premature failure [72].

## 4. Summary and Conclusions

In this study, a combination of high resolution TEM, EDS mapping, and atomistic simulations were used to quantify spatial distribution of dopants and SSRO within amorphous complexions in Cu-rich nanocrystalline alloys. The results presented here allow the following conclusions to be drawn:

(1) Spatial patterning of different dopant species was discovered within the amorphous complexions in multi-component Cu-rich nanocrystalline alloys. Nb and Ti segregate most heavily to the ACTRs, due to the increased density of ordered packing motifs in those zones. In contrast, Zr prefers the complexion interior where the density of disordered packing motifs is highest.

(2) Local chemical composition within an amorphous grain boundary complexion was found to correlate with trends in SSRO, demonstrating that these features should be tunable and opening the door for microstructural engineering within the complexions themselves through alloy design.

(3) Increased chemical complexity was found to result in more pronounced spatial heterogeneities both normal to and within the grain boundary plane. This structural patterning was found to be very sensitive to dopant ratio and total dopant concentration, giving additional pathways for controlling complexion performance.

Overall, this study provides unique insight into dopant partitioning within amorphous grain boundary and interphase complexions, important features that can improve the strength and toughness of nanocrystalline alloys. Chemical partitioning and increased chemical complexity in the grain boundary region are shown to significantly modify the SSRO distribution and could potentially be used to tune damage resistance in the future. Specifically, one may consider an individual dopant's influence on local structural (dis)order to engineer grain boundaries to meet desired properties.

**Acknowledgements**

This research was supported by the U.S. Department of Energy, Office of Science, Basic Energy Sciences, under Award No. DE-SC0025195. The authors acknowledge the use of facilities and instrumentation at the UC Irvine Materials Research Institute (IMRI), which is supported in part by the National Science Foundation through the UC Irvine Materials Research Science and Engineering Center (DMR-2011967). SEM, FIB, and EDS work was performed using instrumentation funded in part by the National Science Foundation Center for Chemistry at the Space-Time Limit (CHE-0802913).

## Supplementary Materials

### Note 1. X-ray diffraction (XRD) for phase identification and phase fraction

Figure S1 presents XRD curves taken from mechanically alloyed, consolidated, and quenched samples of Cu-Zr, Cu-Zr-Nb, and Cu-Zr-Nb-Ti nanocrystalline powders. The primary phase was observed to be a Cu-rich FCC phase, while minor impurity phases were also observed. ZrC, $ZrO_2$, and NbC were found, depending on the alloy, and total impurity phase content was 1.5 vol.% or less.

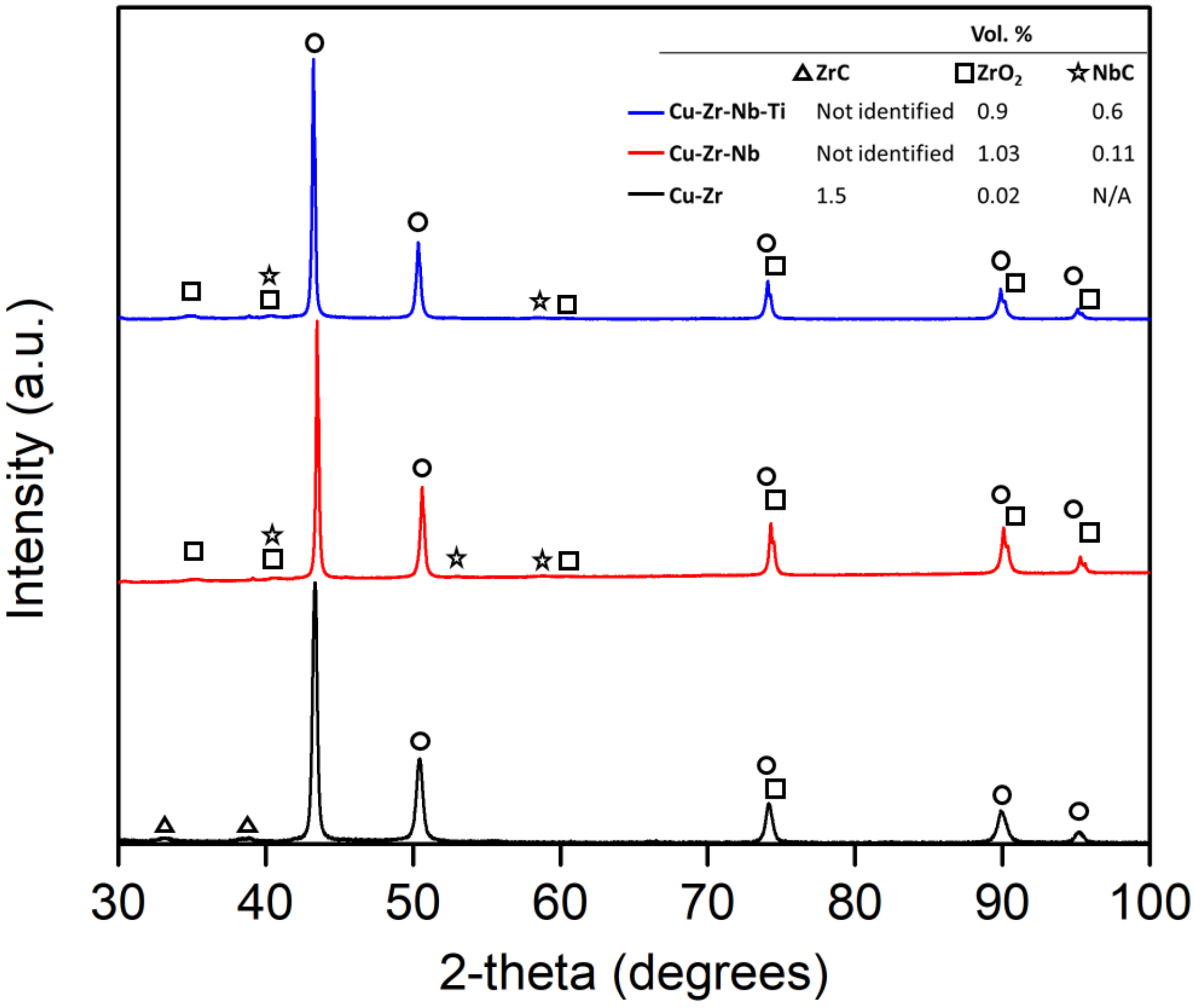

**Figure S1.** XRD scans showing the alloys result in primary FCC Cu (black circles) with small amounts of impurity phases. The total combined volume fraction of carbide or oxide phases was measured to be 1.5% or less.

**Note 2. Qualitative analysis of impurity content in amorphous complexions**

The stearic acid used as a process control agent can contribute O, C, and N to the sample, motivating the study of impurity distribution. Figures S2-S4 show EDS maps from Cu-Zr, Cu-Zr-Nb, and Cu-Zr-Nb-Ti samples, respectively. We note that EDS does not provide reliable quantitative information on light elements such as these impurities, so all discussion remains qualitative in nature here. In all samples, N is distributed uniformly, with no enrichment observed in either the grain interior or grain boundary complexion. C and O show some evidence of segregation to the complexions, with a stronger signal being seen in the grain boundary region than in the grain interior. This could occur due to an affinity for bonding with the segregating metal species (Zr, Nb, and/or Ti) [1] or could come from a tendency to segregate to the grain boundary region themselves [2].

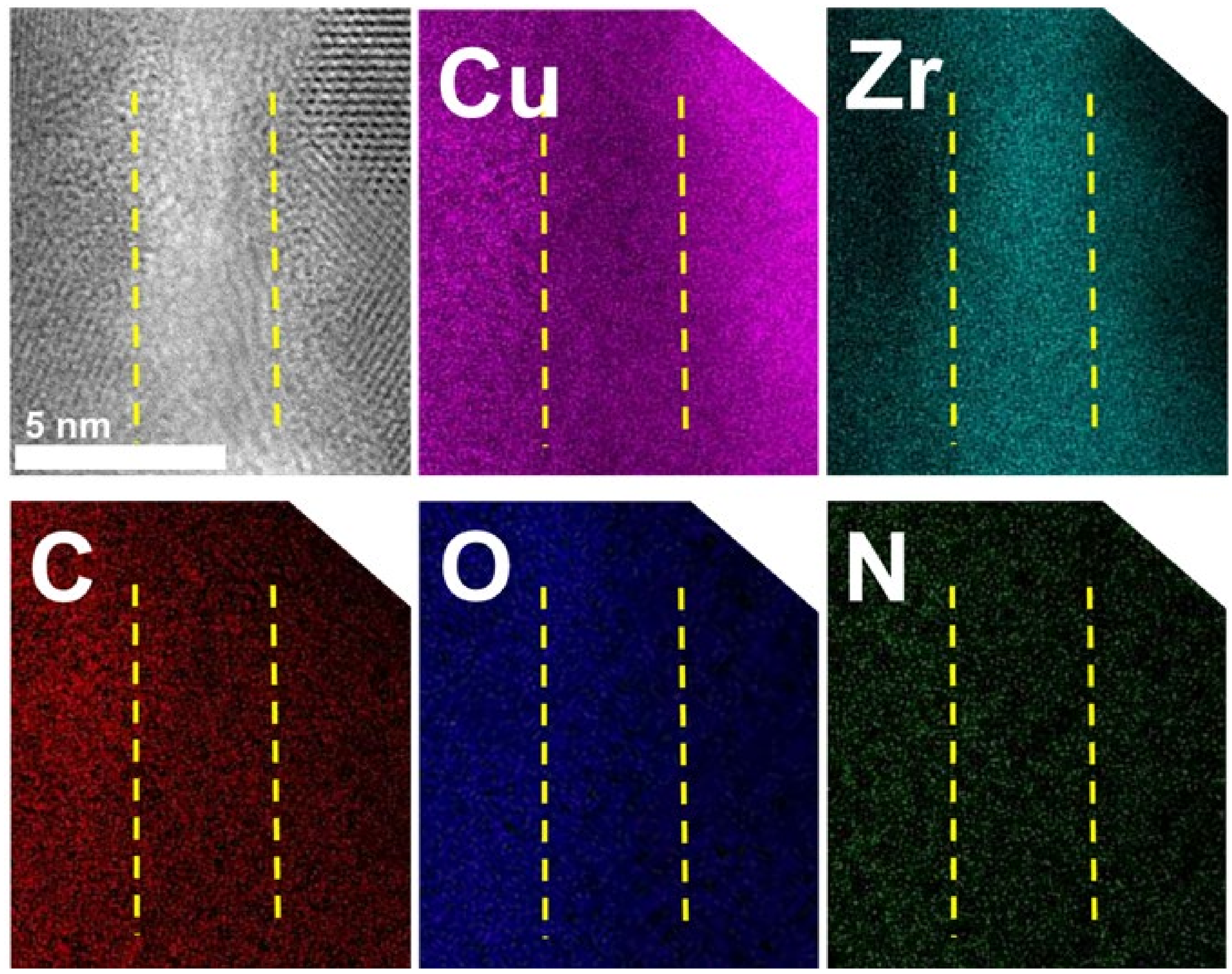

**Figure S2. EDS maps of planned metal dopant and impurity species across an amorphous grain boundary complexions in Cu-Zr.  Zr, O, and C enrichment is observed in the grain boundary regions.**

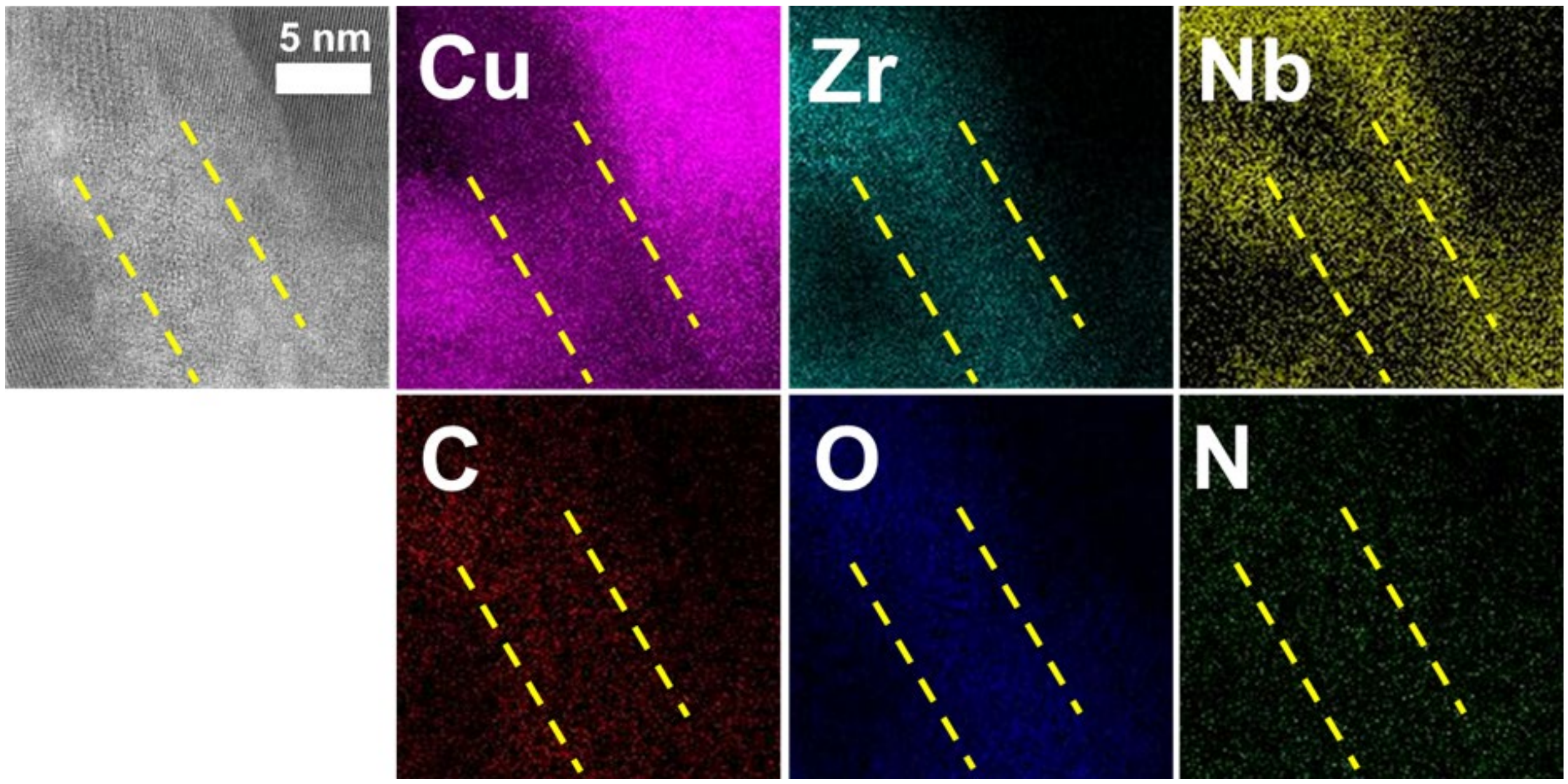

**Figure S3. EDS maps of planned metal dopant and impurity species across an amorphous grain boundary complexions in Cu-Zr-Nb. Zr, Nb, O, and C enrichment is observed in the grain boundary regions.**

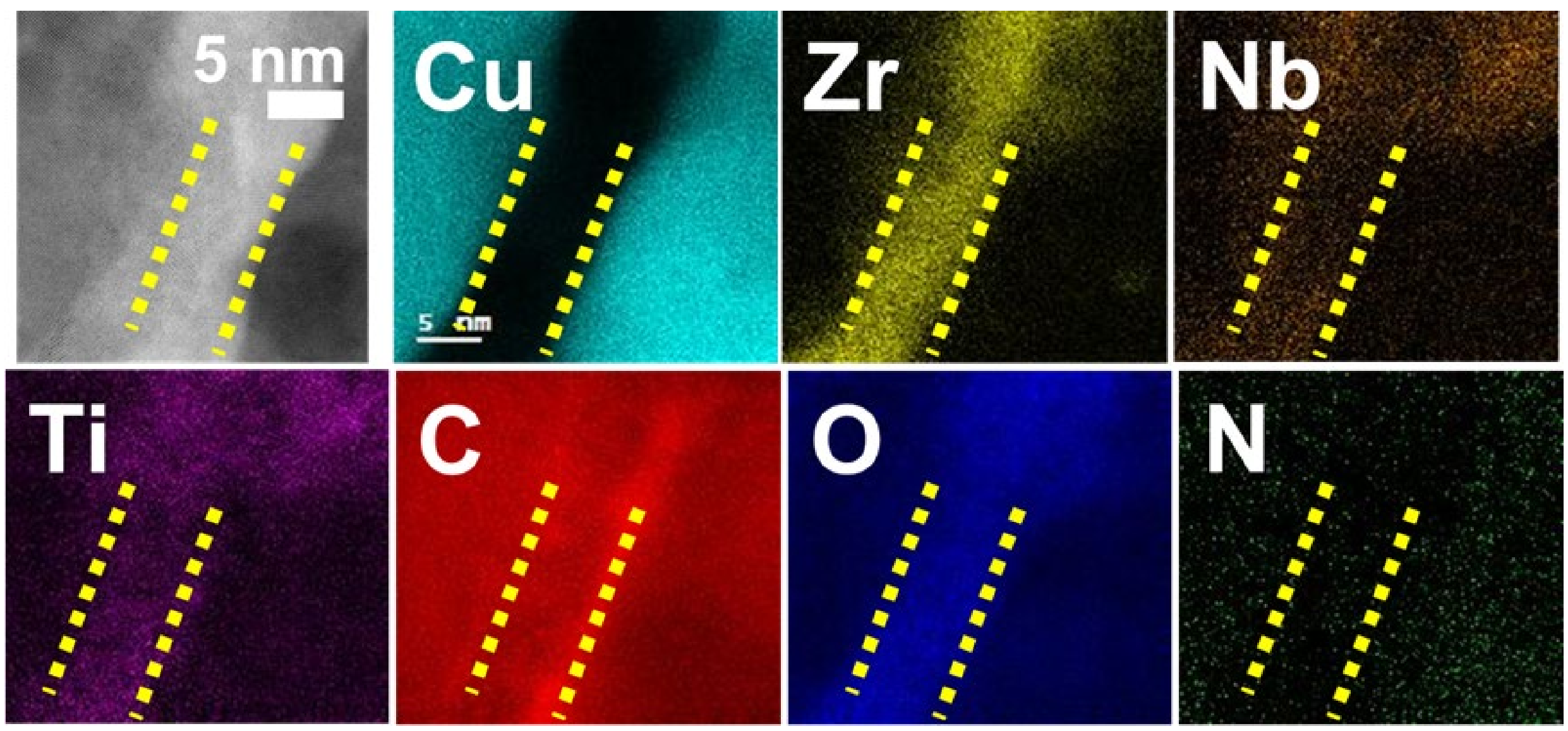

**Figure S4. EDS maps of planned metal dopant and impurity species across an amorphous grain boundary complexions in Cu-Zr-Nb-Ti. Zr, Nb, Ti, O, and C enrichment is observed in the grain boundary regions.**

**Note 3. Nanobeam electron diffraction across an amorphous grain boundary complexion**

Figure S5 presents a set of nanobeam electron diffraction patterns from one amorphous complexion in Cu-Zr, showing the two confining grains, the two ACTRs, and the complexion interior. Several challenges become immediately obvious from these patterns. First, a set of diffraction spots denoted by purple arrows can be seen in all the diffraction patterns in Figure S5, even for the two grains, Grain 1 and Grain 2, which have different crystallographic orientations. This signals that an overlapping grain must be present above or below the features targeted here, complicating attempts at detailed analysis of the different regions. The existence of overlapping grains is also evidenced by the Moiré fringes in the bottom left of Figure S5(b). In addition, the impact of grain boundary curvature out of the viewing plane can be found by tracking the diffraction spots denoted by green arrows in the diffraction patterns in Figure S11. These spots are strongest in Grain 2 and most likely associated with the lattice of that crystal. However, these spots can also be seen in the three patterns in Figures S5(d), (e), and (f) from the complexion which should not have any long-range order. While this set of patterns is imperfect, some qualitative features associated with the complexion can be observed. An amorphous halo marked by blue arrows is observed around the central beam spot, indicative of the disordered structure of the complexion. Unfortunately, the complications associated with overlapping grains and grain boundary curvature degrade the quality of signal that can be obtained and quantitative analysis of this halo was not possible. Extensive sample tilting was attempted to work around these issues and multiple complexions were investigated, yet the very fine grain size of the nanocrystalline alloys studied here precluded direct experimental measurement of SSRO.

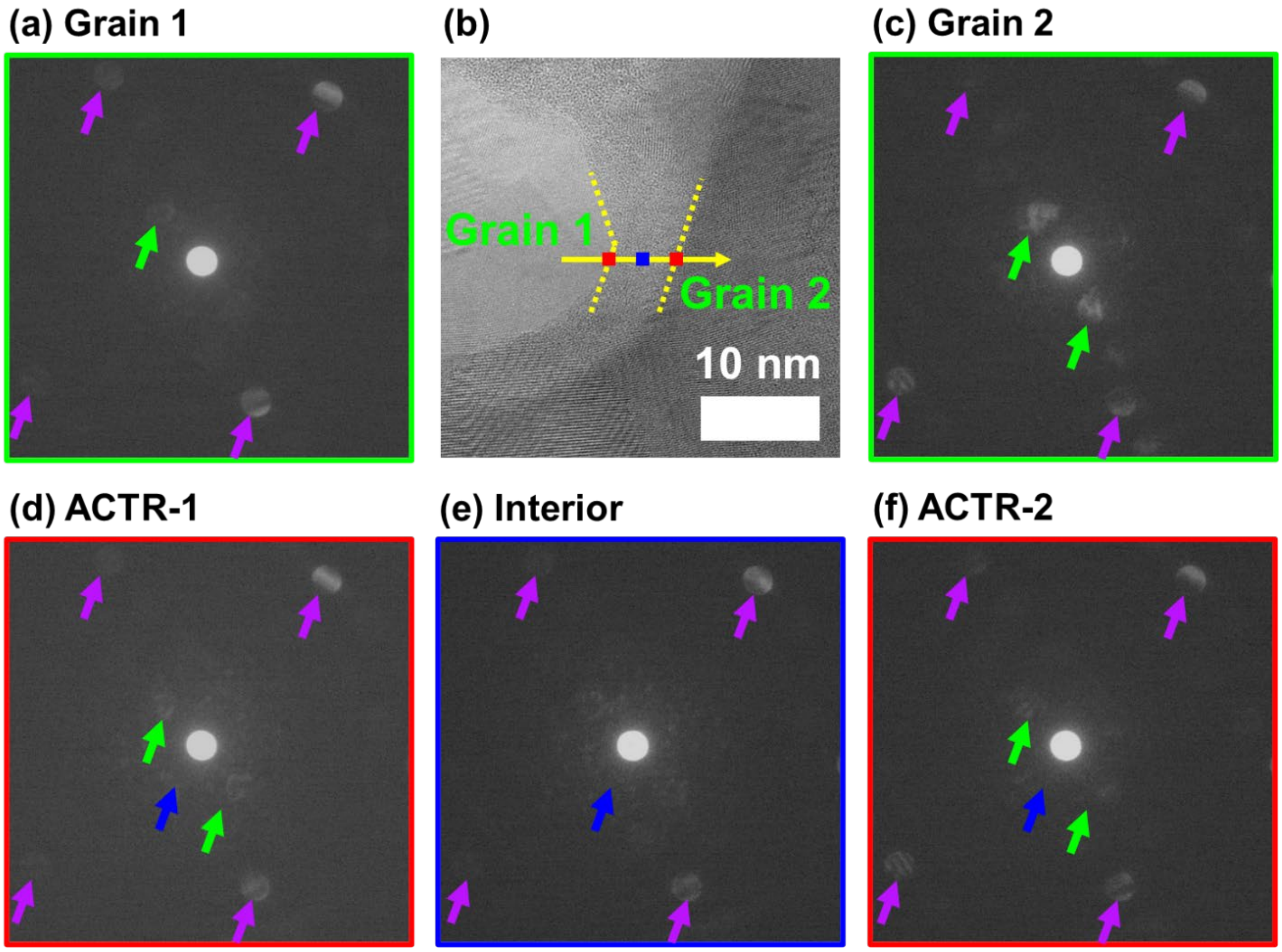

**Figure S5. (a,c) Nanobeam electron diffraction patterns of the grains Grain 1 and Grain 2 abutting the (b) amorphous complexion. Purple arrows denote crystalline spots that appear in all patterns, suggesting an overlapping grain. Green arrows indicate diffraction spots from Grain 2 that are also found in the diffraction patterns (d,e,f) of the complexion, demonstrating that high grain boundary curvature is an issue. Blue arrows indicate the amorphous halo, which is most pronounced for the (e) interior diffraction pattern yet also present in (d,f) the ACTRs.**

## Note 4. Potential energy and gradient data for bi-crystal samples with higher dopant content

Energy convergence for the atomistic simulations was assessed using a multi-scale gradient analysis with four different temporal resolutions. Figures S6-S8 show complete data for potential energy as a function of simulation time. The potential energy gradient was calculated using block-averaging approaches: 0.2 ps blocks for high-frequency analysis, 1.0 ps blocks (five-point averaging) for medium-frequency analysis, 2.0 ps blocks (ten-point averaging) for intermediate-frequency analysis, and 20 ps blocks (hundred-point averaging) for low-frequency analysis. This multi-scale approach enables separation of thermal fluctuations from systematic energy drift. Convergence was determined when energy gradients remained below $10^{-4}$ eV per atom over 20 ps periods across all temporal scales. 120 ps of simulation time is enough to achieve complete convergence.

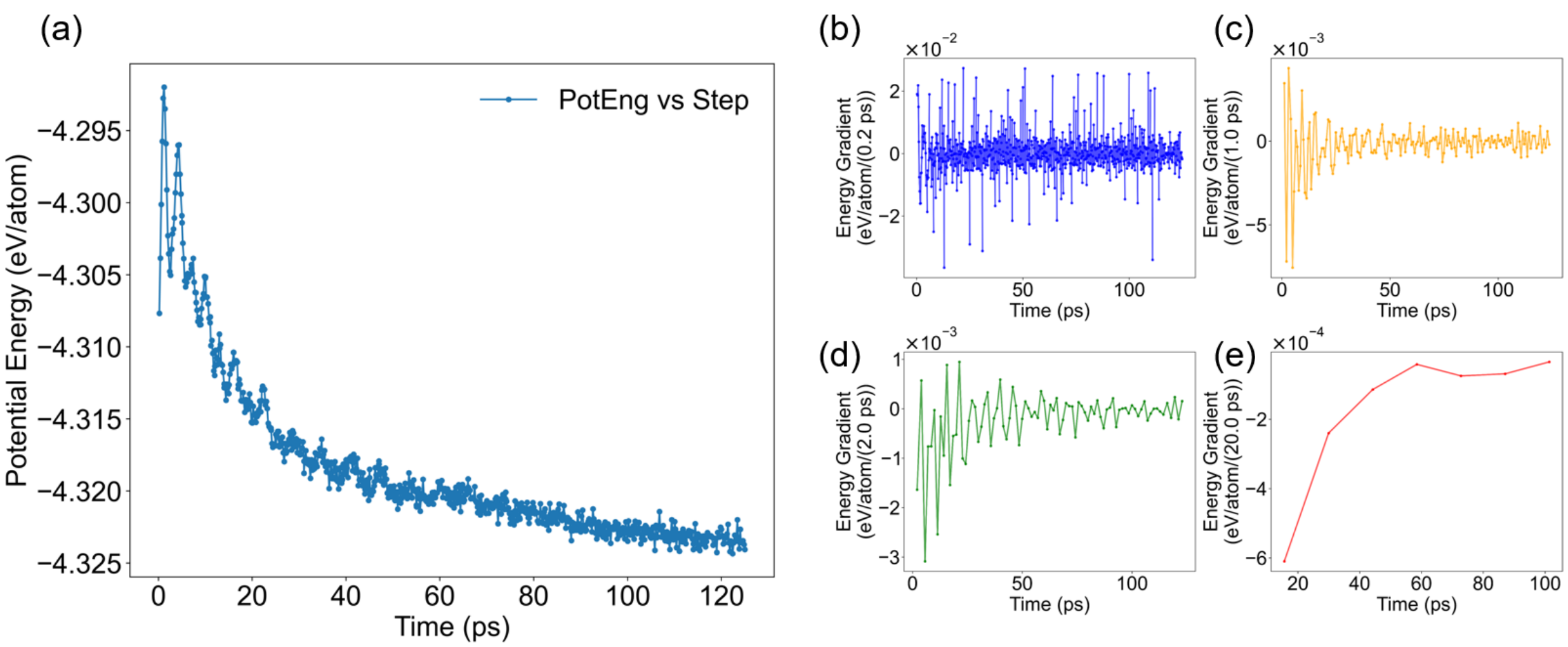

**Figure S6. Complexion with composition of Cu-45Zr potential energy and energy gradient data. (a) Potential energy evolution over ~120 ps. (b-e) Energy gradient analysis at different temporal resolutions: (b) 0.2 ps blocks showing thermal fluctuations, (c) 1.0 ps blocks with five-point averaging, (d) 2.0 ps blocks with ten-point averaging, and (e) 20 ps blocks with hundred-point averaging.**

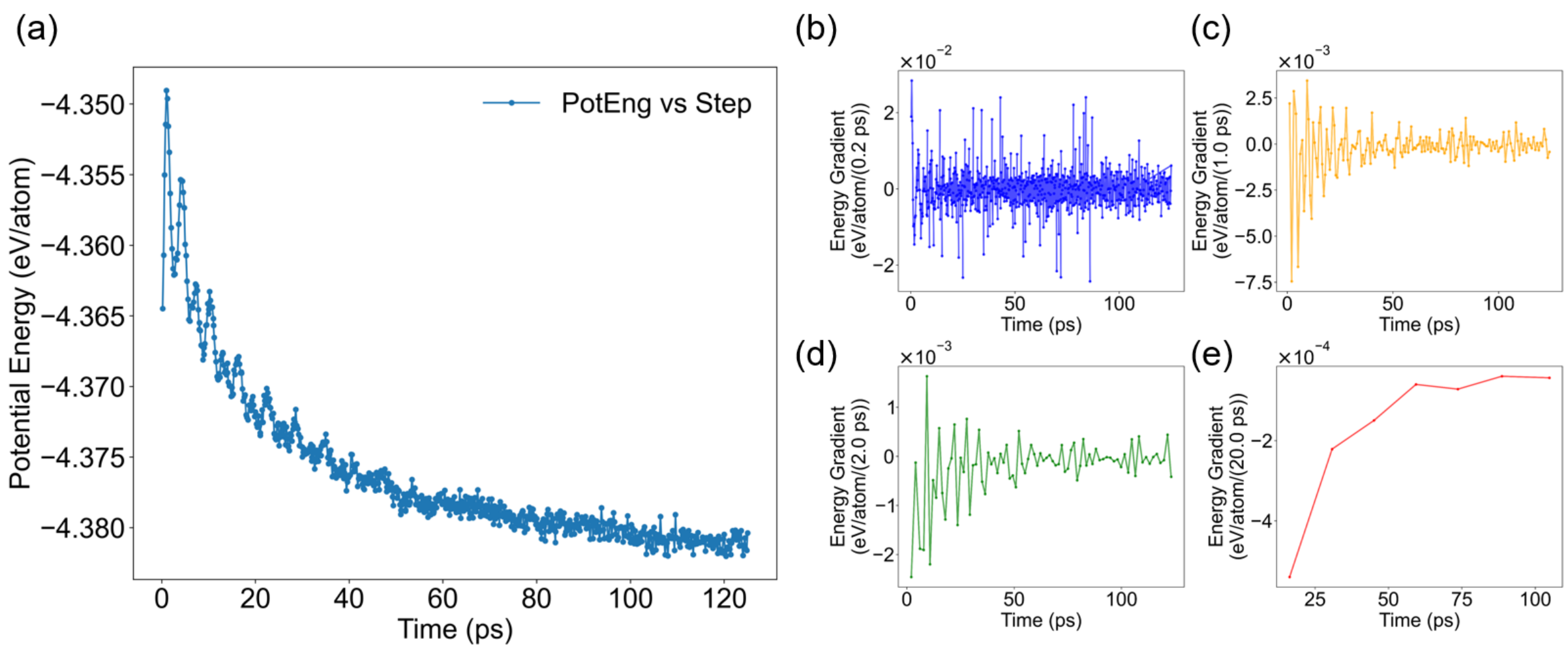

**Figure S7. Complexion with composition of Cu-40Zr-5Nb potential energy and energy gradient data. (a) Potential energy evolution over ~120 ps. (b-e) Energy gradient analysis at different temporal resolutions: (b) 0.2 ps blocks showing thermal fluctuations, (c) 1.0 ps blocks with five-point averaging, (d) 2.0 ps blocks with ten-point averaging, and (e) 20 ps blocks with hundred-point averaging.**

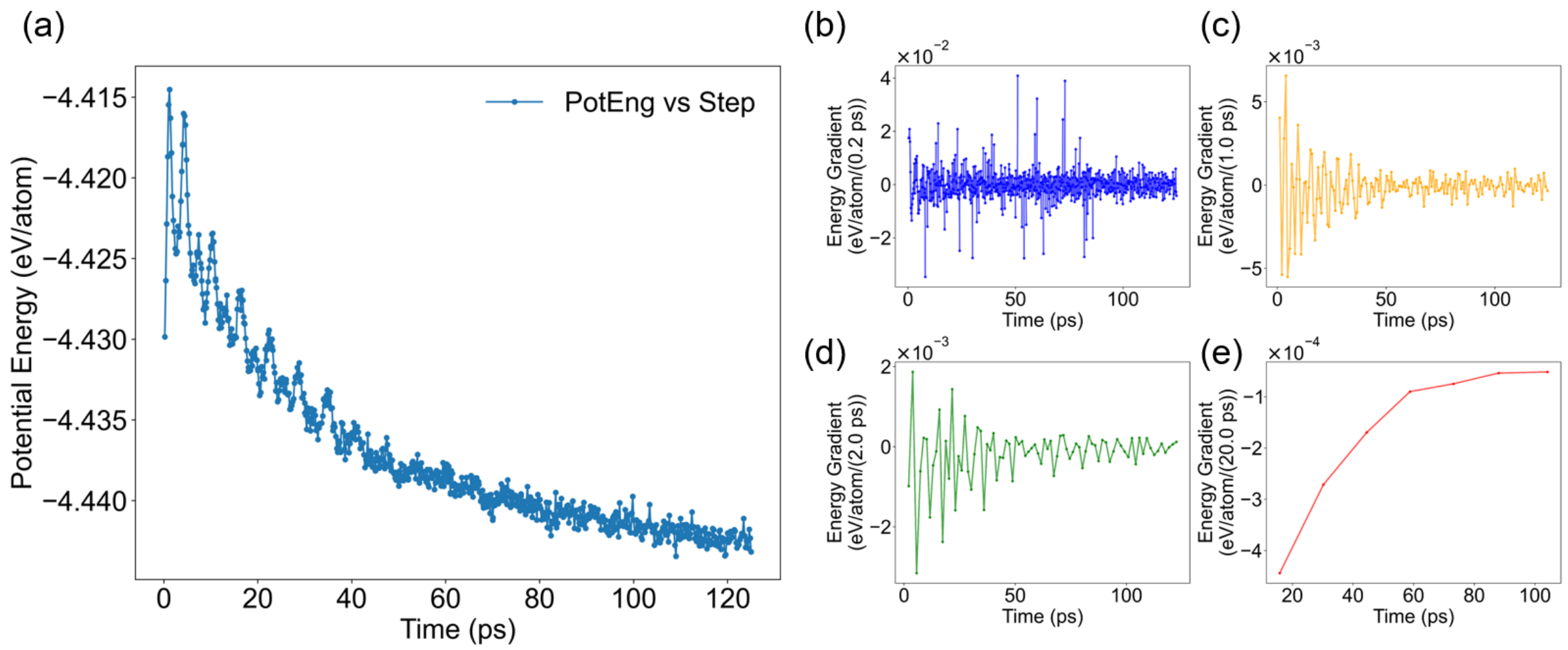

**Figure S8. GB complexion with composition of Cu-38Zr-7Nb potential energy and energy gradient data. (a) Potential energy evolution over ~120 ps. (b-e) Energy gradient analysis at different temporal resolutions: (b) 0.2 ps blocks showing thermal fluctuations, (c) 1.0 ps blocks with five-point averaging, (d) 2.0 ps blocks with ten-point averaging, and (e) 20 ps blocks with hundred-point averaging.**

**Note 5. Calculation of the disorder parameter**

The bond-orientational disorder parameter is calculated for atoms in bi-crystal samples and small metallic samples [3], similar with previous work [4], which measures the similarity between each atom's local bonding environment to that within a cutoff by method outlined by von Alfthan et. al. [5]. In this study, a cutoff radius of 3.0 Å was employed to ensure that atoms in the ACTR region had no neighboring atoms from the FCC phase within this distance. The local structure around particle *i* is defined as a set of complex numbers:

$$\bar{Y}_m^l(i) = \frac{1}{N_i}\sum_{j\in N_i} Y_m^l\left(\theta_{ij}, \Phi_{ij}\right)$$

where $N_i$ is the number of neighbor atoms, $\theta_{ij}$ is the colatitude and $\Phi_{ij}$ is the azimuthal angle betweens atoms *i* and *j*. The normalized complex vector is calculated by:

$$\hat{Y}_m^l = \frac{\bar{Y}_m^l}{\left|\bar{Y}_m^l\right|}$$

where $l = 6$. This calculation is performed in LAMMPS by using the *orientorder/atom* compute and setting the *components* value to 6. The outcome of this step is a $2\times(2l + 1)$ component array corresponding to the real and imaginary parts of normalized $\hat{Y}_m^l$. The similarity between the angular distributions of a central atom *i* and its neighboring atoms can be evaluated as:

$$s_{ij} = \sum_{m=-l}^{l} \hat{Y}_m^l(i)\hat{Y}_m^{l*}(j)$$

Then, the disorder parameter for each atom is calculated:

$$d_i = 1 - \frac{s_{ij}}{N_i}$$

Where the higher value means more disorder and value 0 correspond to atoms in perfect crystal [6].

**Note 6. Database for ML-IAP**

As illustrated in Figures S9-S12, Materials Project [7] and the Thermo-calc [8] were used to help establish the bottom of formation energy of the system. All the structures used by the Materials Project could be found in Table S1. Three compositions: $Cu_{55}Zr_{30}Nb_{15}$, $Cu_{35}Zr_{15}Nb_{50}$ and CuZrNb are used to create phase diagrams via Thermo-calc, where $Zr_3Cu_8$, $Zr_7Cu_{10}$, $Zr_{14}Cu_{51}$ and BCC B2 phase $Nb_{95}Zr_3Cu_2$ were selected.

Besides the stable configurations from phase diagrams, FCC, BCC and HCP solid solutions are created to cover as much local environment as possible. 14 compositions were selected for FCC, and only equal atomic composition was selected for BCC and HCP structures as shown in Table S2. To cover more local environment, 5 different atomic configurations were selected for each of the compositions. This is done by combining random structure generation with graph-based clustering to find diverse atomic configurations for density functional theory calculations. The process begins by generating 5000 random atomic configurations for each target composition within FCC, BCC, or HCP crystal lattice frameworks, where atomic species are randomly placed on lattice sites according to the stoichiometry. Each generated structure is then converted into a graph representation with atoms as nodes and interatomic connections within a 3.0 Å cutoff distance as edges. Structural fingerprints are calculated to characterize local atomic environments through multiple descriptors: local composition features capturing the average and variance of element composition within the first coordination shell of each atom type, pairwise distance statistics including mean, standard deviation, and coordination numbers for each element pair, and geometric descriptors of local coordination patterns. To identify structurally different configurations from this large set, hierarchical clustering is used to group similar structures. Representative structures are selected from each cluster as those closest to cluster centers. This method ensures complete sampling of local atomic environments while reducing computational redundancy by removing structurally similar configurations, providing an optimal training set that maximizes structural diversity for machine learning potential development while minimizing DFT computational cost.

In order to cover more configurations at higher temperatures including crystal and amorphous structures, ab initio molecular dynamics is used within VASP. Each of the 0 K structures is equilibrated at 800 K and 1500 K, then 10 frames are randomly extracted to capture atomic local environments at high temperature. Also, to capture amorphous structures especially the amorphous grain boundary, each of the structures was heated to 5000 K and quenched.

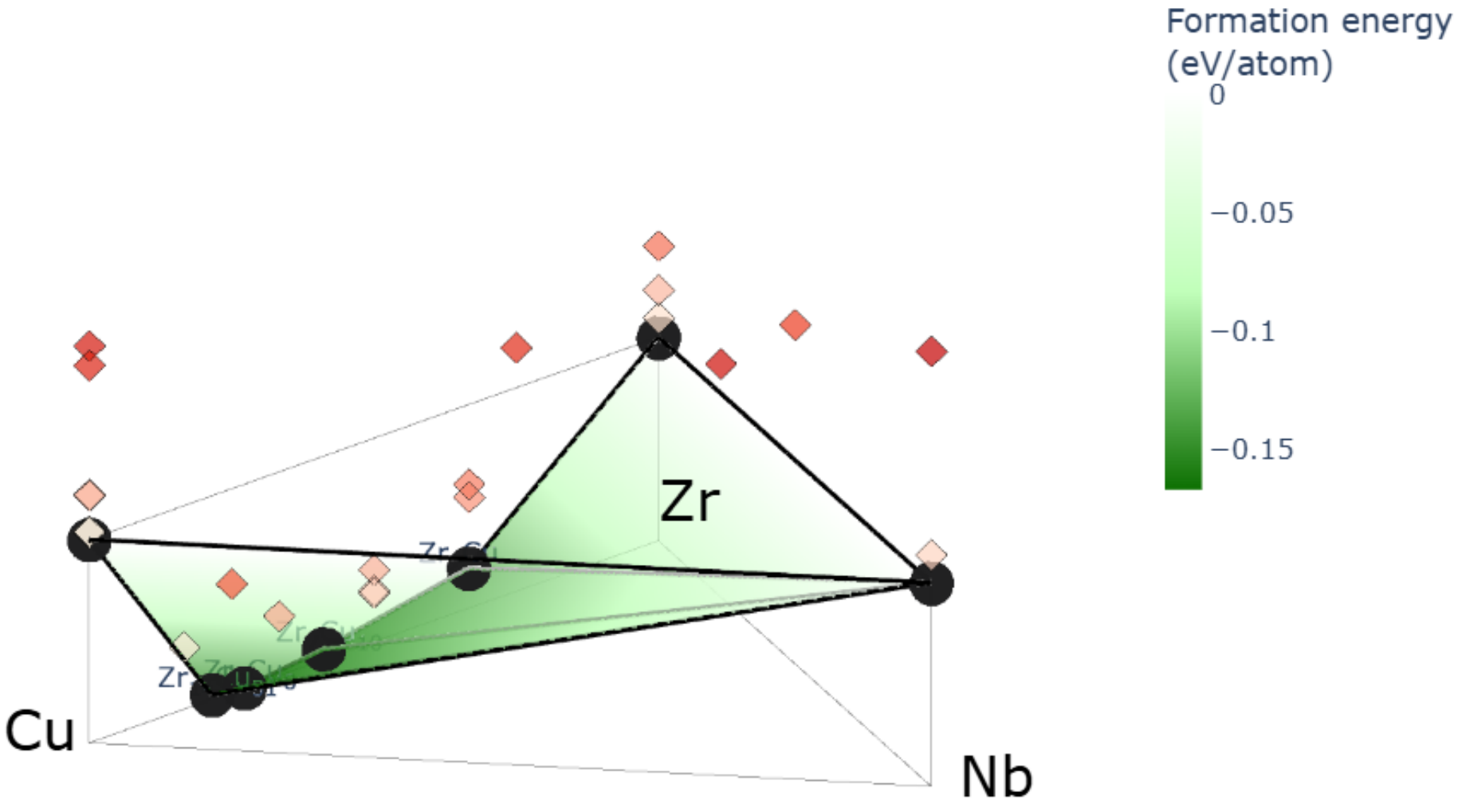

**Figure S9. Ternary phase diagram of Cu-Zr-Nb system from Materials Project.**

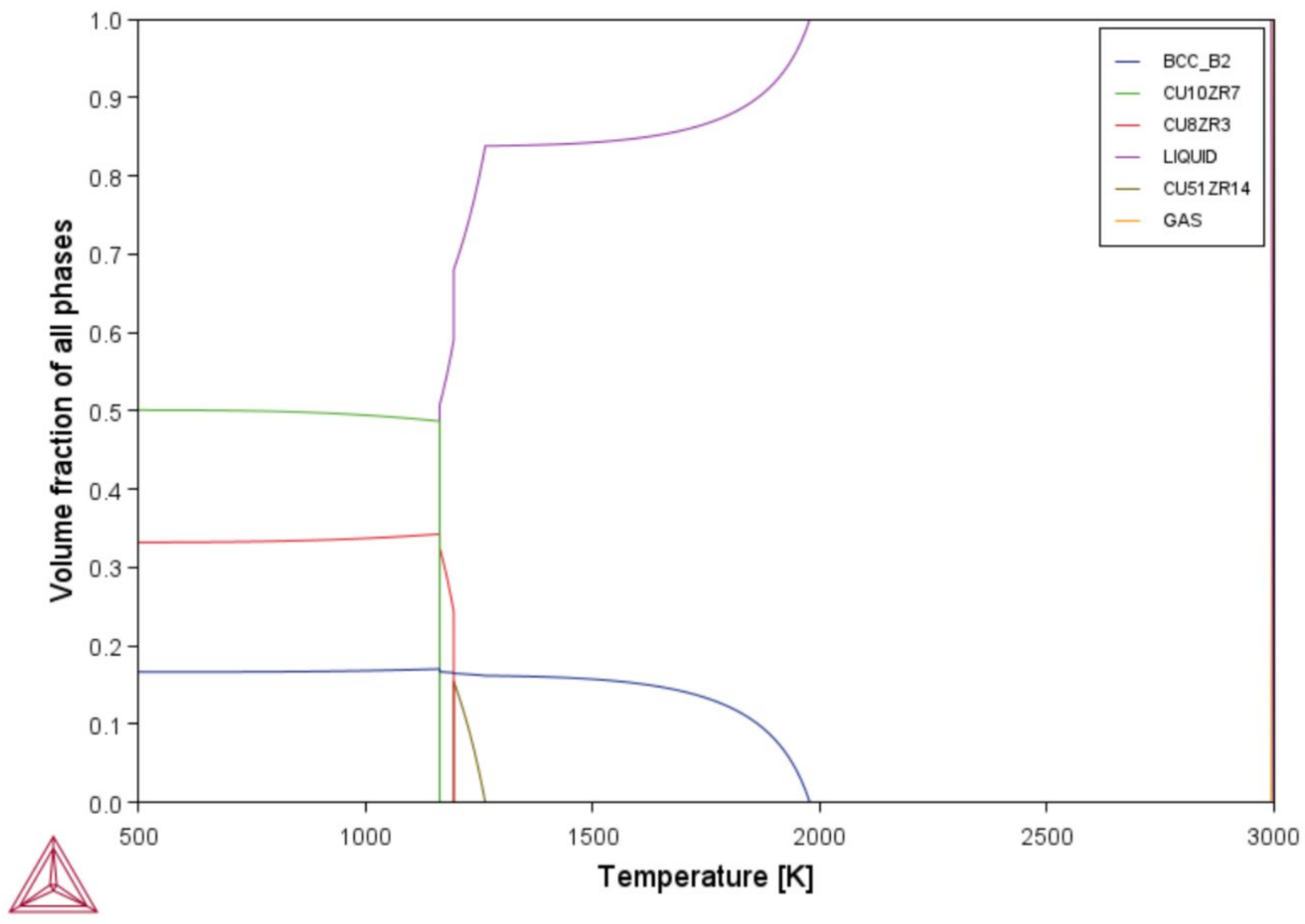

**Figure S10. Phase diagram of $Cu_{55}Zr_{30}Nb_{15}$ from Thermo-calc.**

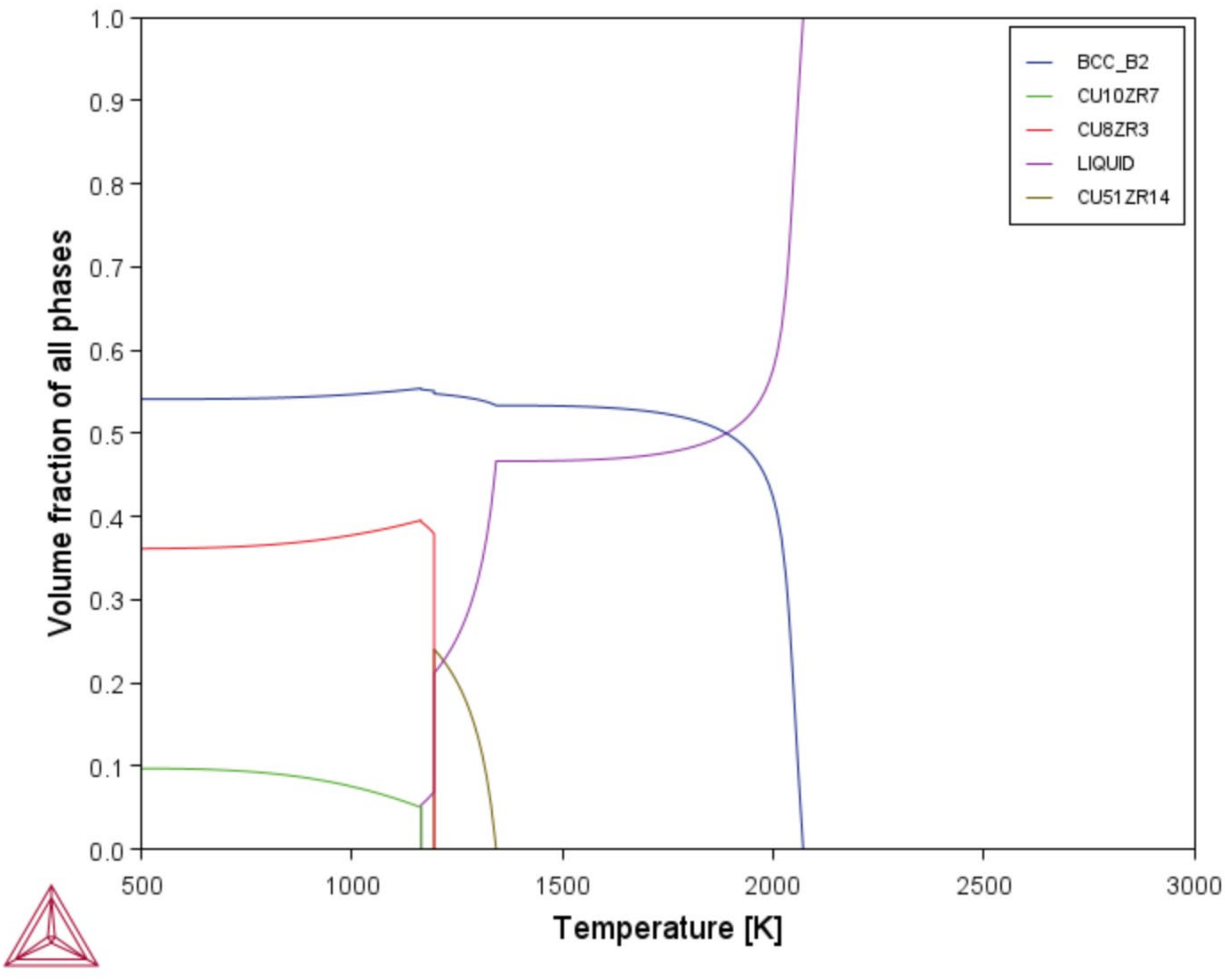

**Figure S11. Phase diagram of $Cu_{35}Zr_{15}Nb_{50}$ from Thermo-calc.**

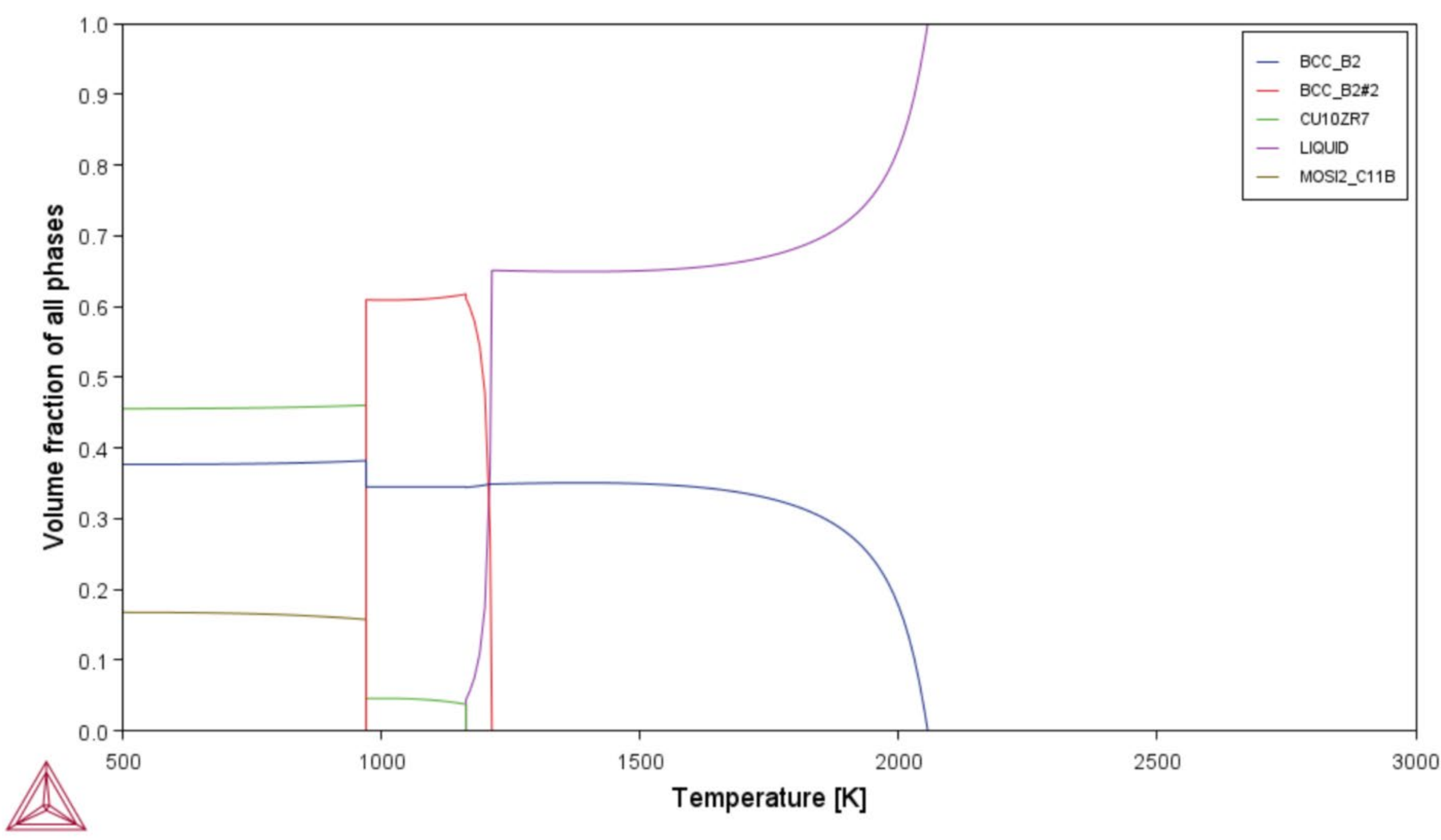

**Figure S12. Phase diagram of $Cu_{33.3}Zr_{33.3}Nb_{33.3}$ from Thermo-calc.**

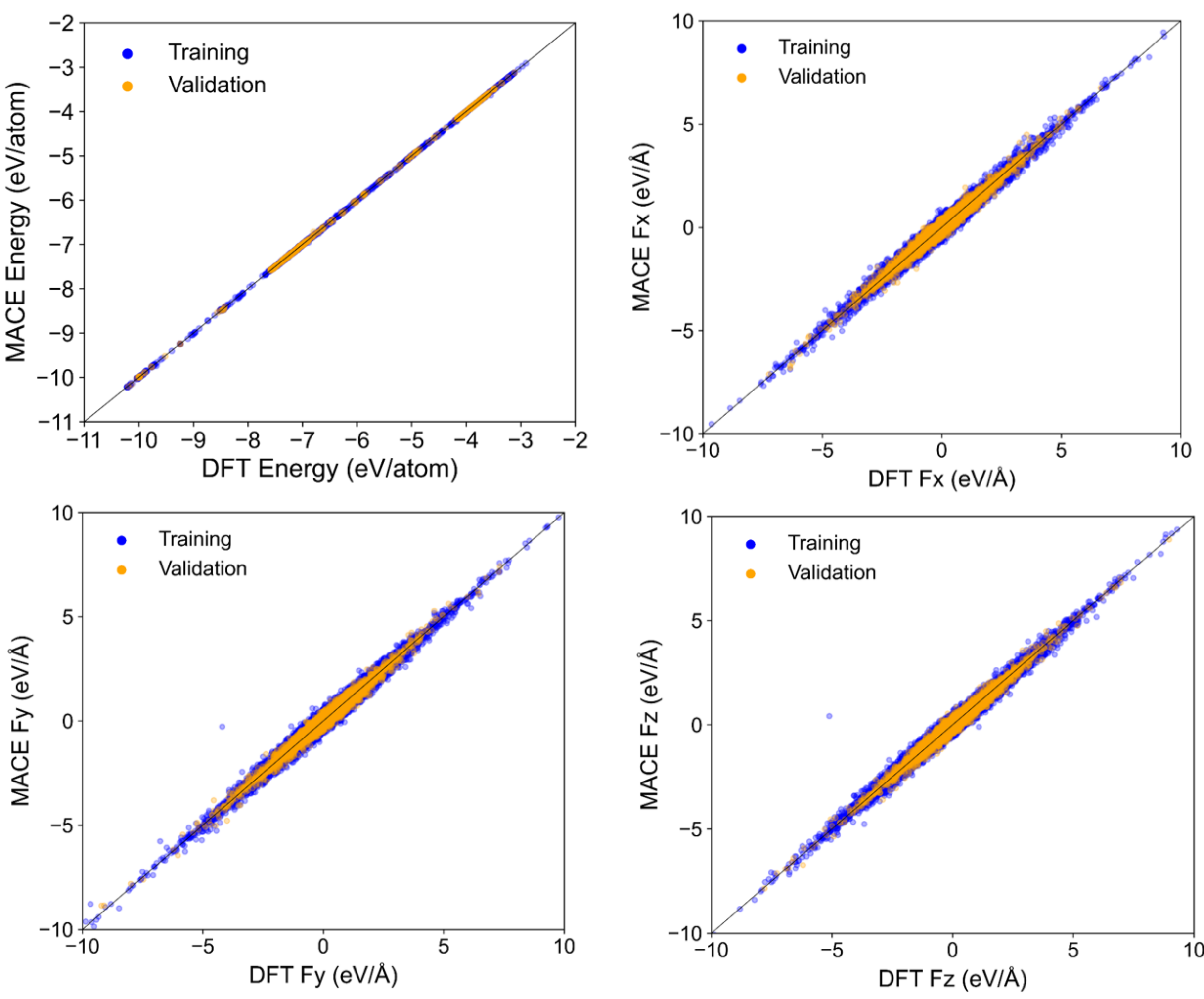

**Figure S13. ML potential training results of energy and force on root mean square errors.**

**Table S1. Structure used from Materials Project.**

| Structure name | Materials Project code |
| --- | --- |
| $Zr_3Cu_8$ | Mp-1195821 |
| $Zr_7Cu_{10}$ | Mp-1188077 |
| $Zr_{14}Cu_{51}$ | Mp-1216441 |
| Cu | Mp-1010136 |
| Cu | Mp-1056079 |
| Cu | Mp-1059259 |
| Cu | Mp-1120774 |
| Cu | Mp-30 |
| Cu | Mp-989695 |
| Cu | Mp-989782 |
| Cu | Mp-998890 |
| $Nb_3Cu$ | Mp-1186214 |
| Nb | Mp-1094120 |
| Nb | Mp-1104341 |
| Nb | Mp-2647103 |
| Nb | Mp-2739273 |
| Nb | Mp-75 |
| Nb | Mp-8636 |
| $Zr_2Cu$ | Mp-1077372 |

| $Zr_2Cu$ | Mp-193 |
| --- | --- |
| $Zr_2Cu$ | Mp-583800 |
| $Zr_3Cu$ | Mp-580287 |
| $ZrCu_2$ | Mp-1072655 |
| $ZrCu_3$ | Mp-1188040 |
| $ZrCu_5$ | Mp-30603 |
| ZrCu | Mp-1067210 |
| ZrCu | Mp-1080022 |
| ZrCu | Mp-2018976 |
| ZrCu | Mp-2210 |
| Zr | Mp-1056376 |
| Zr | Mp-1077723 |
| Zr | Mp-1178608 |
| Zr | Mp-131 |
| Zr | Mp-41 |
| Zr | Mp-8635 |
| ZrNb | Mp-1215202 |

**Table S2. Composition used in solid solution**

| Structure | Composition |
|---|---|
| FCC | $Cu_{16}Zr_{16}Nb_{16}$ |
| FCC | $Cu_{8}Zr_{8}Nb_{32}$ |
| FCC | $Cu_{8}Zr_{32}Nb_{8}$ |
| FCC | $Cu_{32}Zr_{8}Nb_{8}$ |
| FCC | $Cu_{8}Zr_{16}Nb_{24}$ |
| FCC | $Cu_{8}Zr_{24}Nb_{16}$ |
| FCC | $Cu_{16}Zr_{8}Nb_{24}$ |
| FCC | $Cu_{24}Zr_{8}Nb_{16}$ |
| FCC | $Cu_{16}Zr_{24}Nb_{8}$ |
| FCC | $Cu_{24}Zr_{16}Nb_{8}$ |
| FCC | $Cu_{47}Zr_{1}Nb_{0}$ |
| FCC | $Cu_{47}Zr_{0}Nb_{1}$ |
| FCC | $Cu_{46}Zr_{1}Nb_{1}$ |
| FCC | $Cu_{44}Zr_{2}Nb_{2}$ |
| BCC | $Cu_{18}Zr_{18}Nb_{18}$ |
| HCP | $Cu_{22}Zr_{21}Nb_{21}$ |

**Note 7. Analysis and data for bi-crystal samples with lower dopant content**

**7.1 Potential energy and gradient data for bi-crystal samples**

Energy convergence was assessed using a multi-scale gradient analysis with four different temporal resolutions. Figures S14-S17 show complete data for potential energy as a function of simulation time. The potential energy gradient was calculated using block-averaging approaches: 0.2 ps blocks for high-frequency analysis, 1.0 ps blocks (five-point averaging) for medium-frequency analysis, 2.0 ps blocks (ten-point averaging) for intermediate-frequency analysis, and 20 ps blocks (hundred-point averaging) for low-frequency analysis. This multi-scale approach enables separation of thermal fluctuations from systematic energy drift. Convergence was determined when energy gradients remained below $10^{-4}$ eV per atom over 20 ps periods across all temporal scales. 200 ps of simulation time is enough to achieve complete convergence.

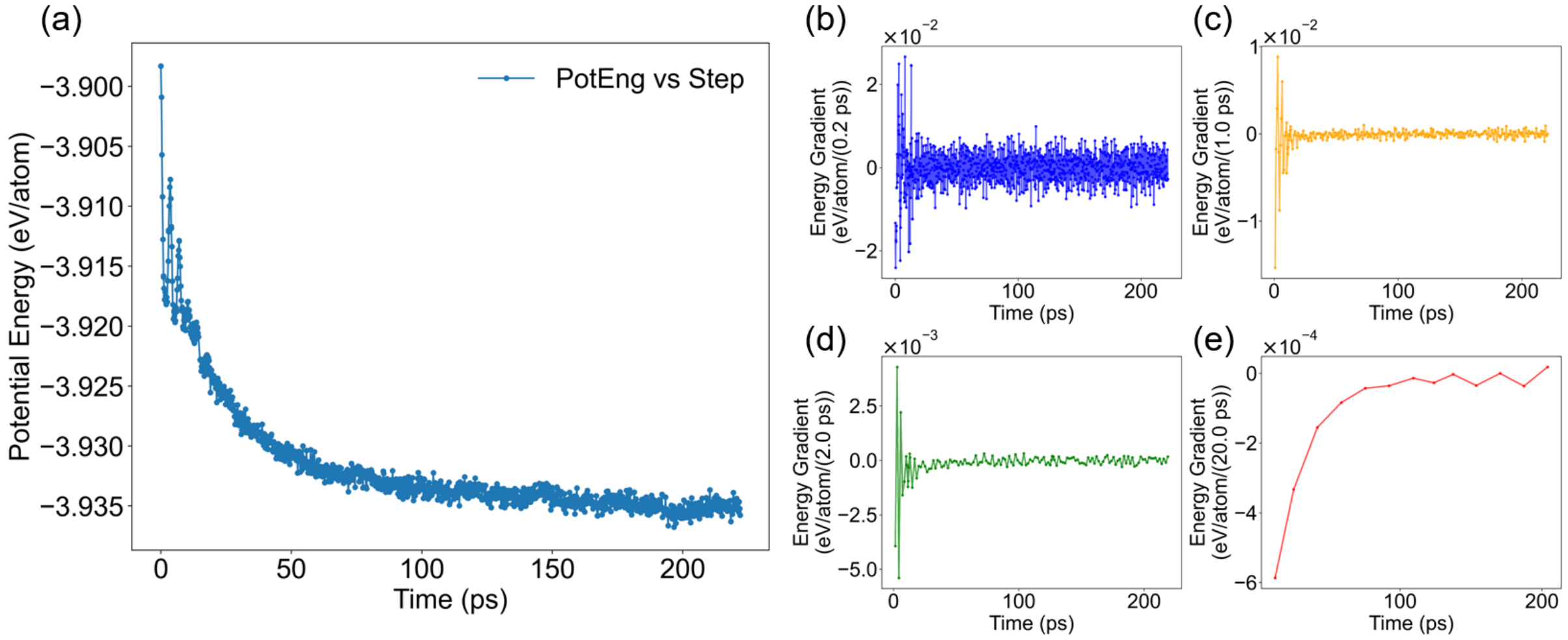

**Figure S14. Complexion with composition of Cu-6Zr-12Nb potential energy and energy gradient data. (a) Potential energy evolution over ~200 ps. (b-e) Energy gradient analysis at different temporal resolutions: (b) 0.2 ps blocks showing thermal fluctuations, (c) 1.0 ps blocks with five-point averaging, (d) 2.0 ps blocks with ten-point averaging, and (e) 20 ps blocks with hundred-point averaging.**

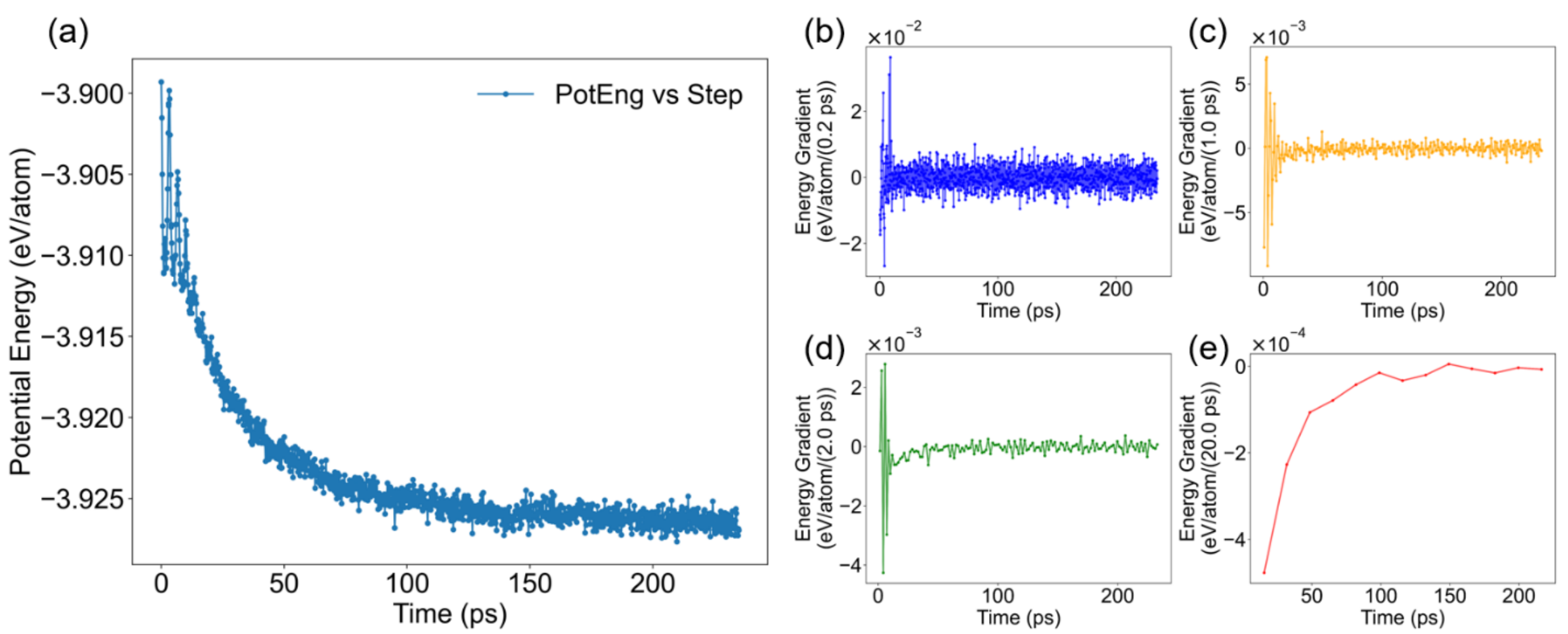

**Figure S15. Complexion with composition of Cu-9Zr-9Nb potential energy and energy gradient data. (a) Potential energy evolution over ~200 ps. (b-e) Energy gradient analysis at different temporal resolutions: (b) 0.2 ps blocks showing thermal fluctuations, (c) 1.0 ps blocks with five-point averaging, (d) 2.0 ps blocks with ten-point averaging, and (e) 20 ps blocks with hundred-point averaging.**

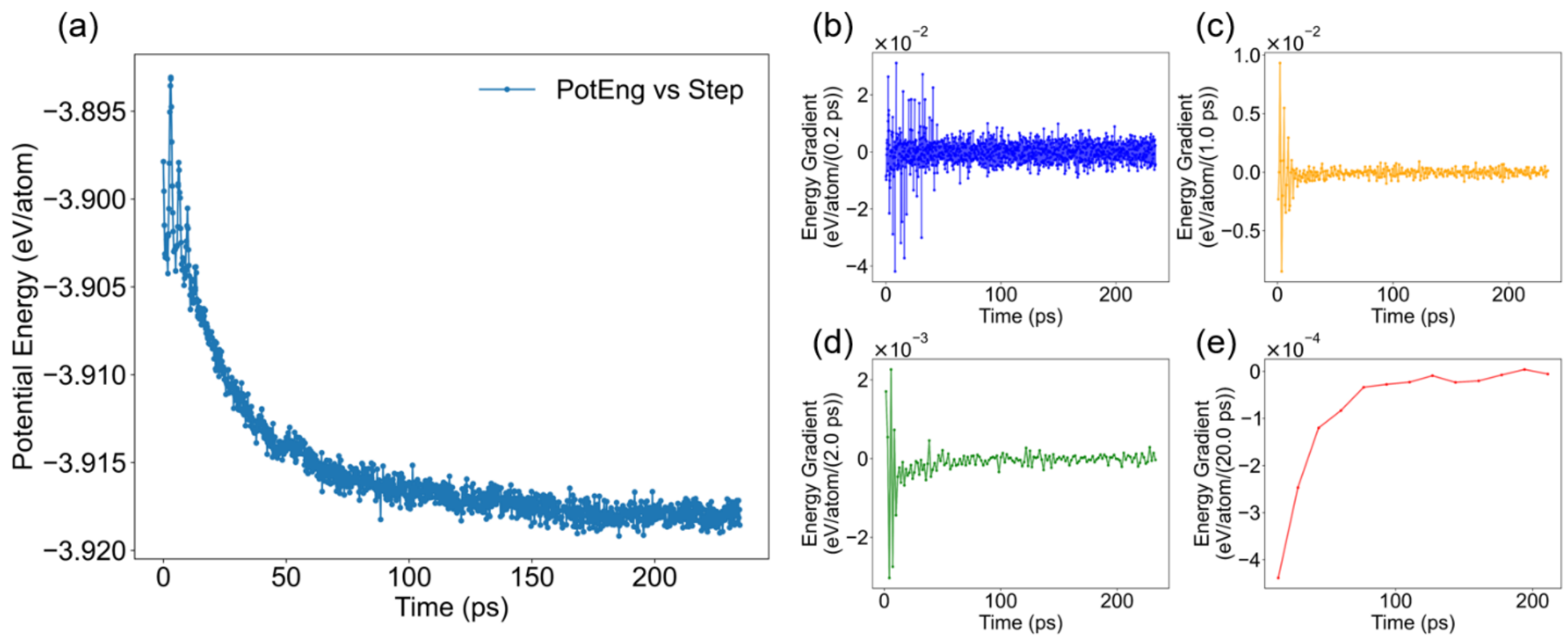

**Figure S16. Complexion with composition of Cu-12Zr-6Nb potential energy and energy gradient data. (a) Potential energy evolution over ~200 ps. (b-e) Energy gradient analysis at different temporal resolutions: (b) 0.2 ps blocks showing thermal fluctuations, (c) 1.0 ps blocks with five-point averaging, (d) 2.0 ps blocks with ten-point averaging, and (e) 20 ps blocks with hundred-point averaging.**

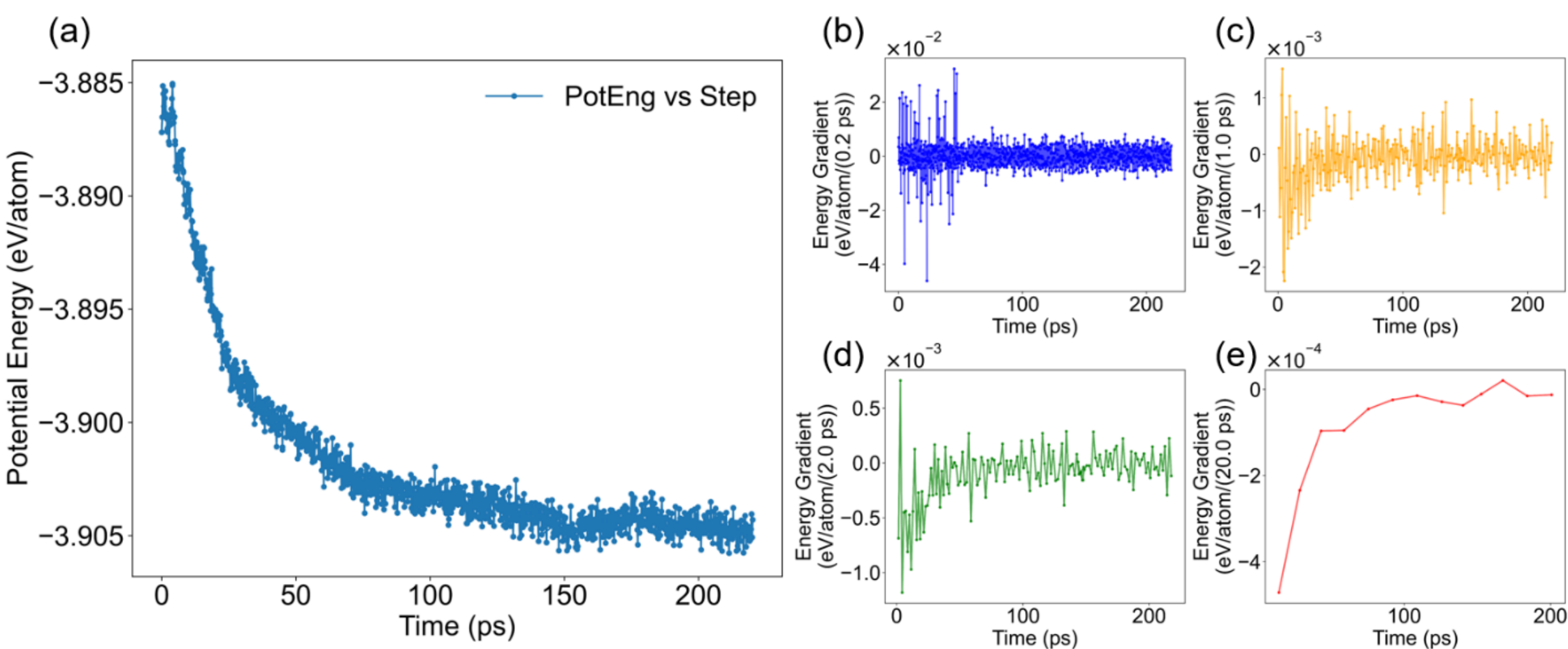

**Figure S17. Complexion with composition of Cu-18Zr potential energy and energy gradient data. (a) Potential energy evolution over ~200 ps. (b-e) Energy gradient analysis at different temporal resolutions: (b) 0.2 ps blocks showing thermal fluctuations, (c) 1.0 ps blocks with five-point averaging, (d) 2.0 ps blocks with ten-point averaging, and (e) 20 ps blocks with hundred-point averaging.**

### 7.2 Disorder parameter analysis and discussion

All of the complexions are depleted of Cu and enriched with the dopant species, as shown in the top row of Figure S18. The bottom row of Figure S18 provides a detailed view of the dopant concentrations across the complexion thickness, where the ACTRs are labeled in grey. For all of the alloys, the Zr concentration is always higher in the complexion interior than in the ACTRs. Looking specifically at the multi-component Cu-Zr-Nb alloys, the Nb concentration always shows a peak value within the ACTRs, with slightly lower but still enriched levels in the complexion interior.

Figure S19 presents measurements of the local disorder parameter for each of the four alloys studied here, with measurements taken in both the ACTRs and complexion interior. In all cases, the complexion interior is more disordered (i.e., has a higher disorder parameter) than the ACTRs, with this effect being relatively small in the Cu-Zr and much more significant in the Cu-Zr-Nb alloys. An asymmetry can be observed in Figure S19 between ACTR-1 and ACTR-2, or the ACTR touching Grain 1 and Grain 2, respectively. For all of the alloys, ACTR-2 has a higher disorder parameter than ACTR-1, although again the magnitude of the asymmetry is material-dependent. Using complexion composition as sample name, it is quite small for Cu-18Zr and Cu-6Zr-12Nb and relatively for Cu-12Zr-6Nb and Cu-9Zr-9Nb.

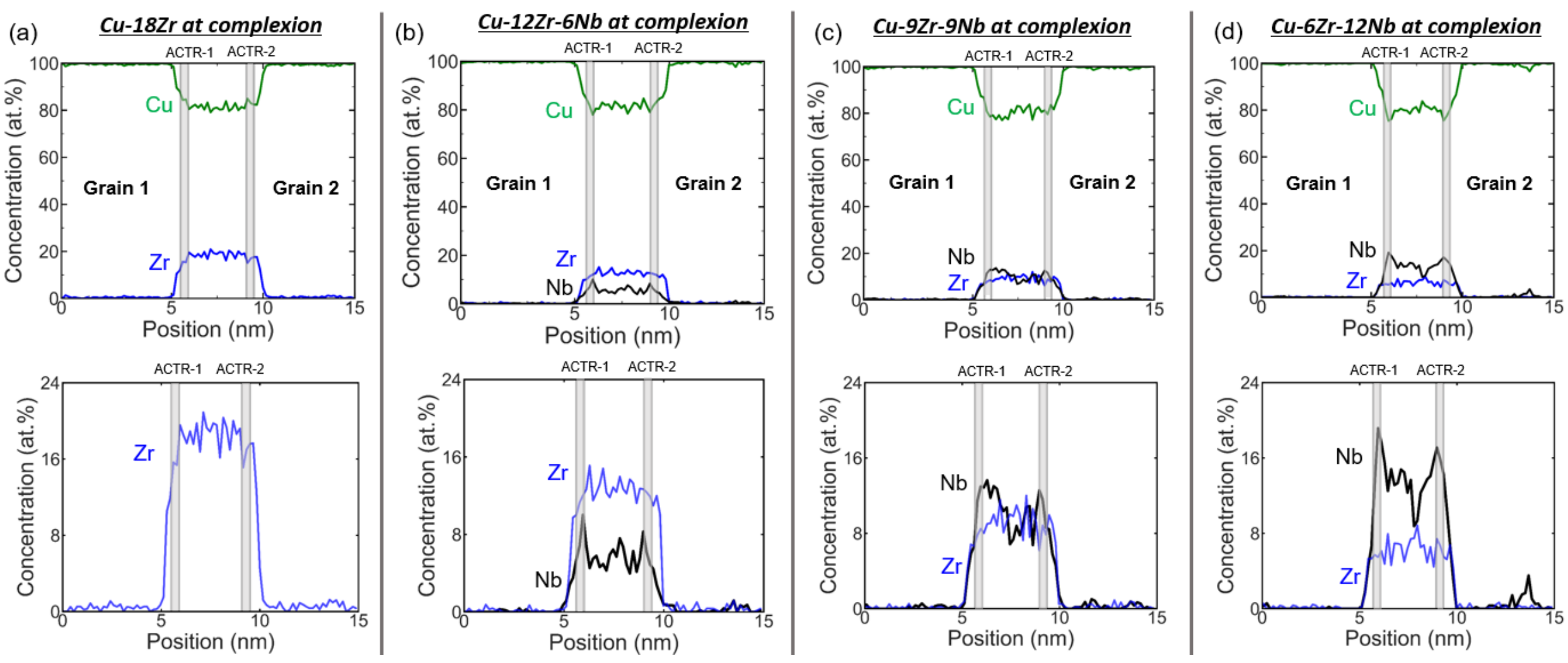

**Figure S18. Chemical concentration maps across an amorphous complexion with composition of (a) Cu-18Zr, (b) Cu-12Zr-6Nb, (c) Cu-9Zr-9Nb, and (d) Cu-6Zr-12Nb. The dopants are enriched in the complexion region for all alloys. Zr concentration is highest in the complexion interior while Nb concentration is highest in the ACTRs (denoted by grey shading), matching the experimental observations of this study.**

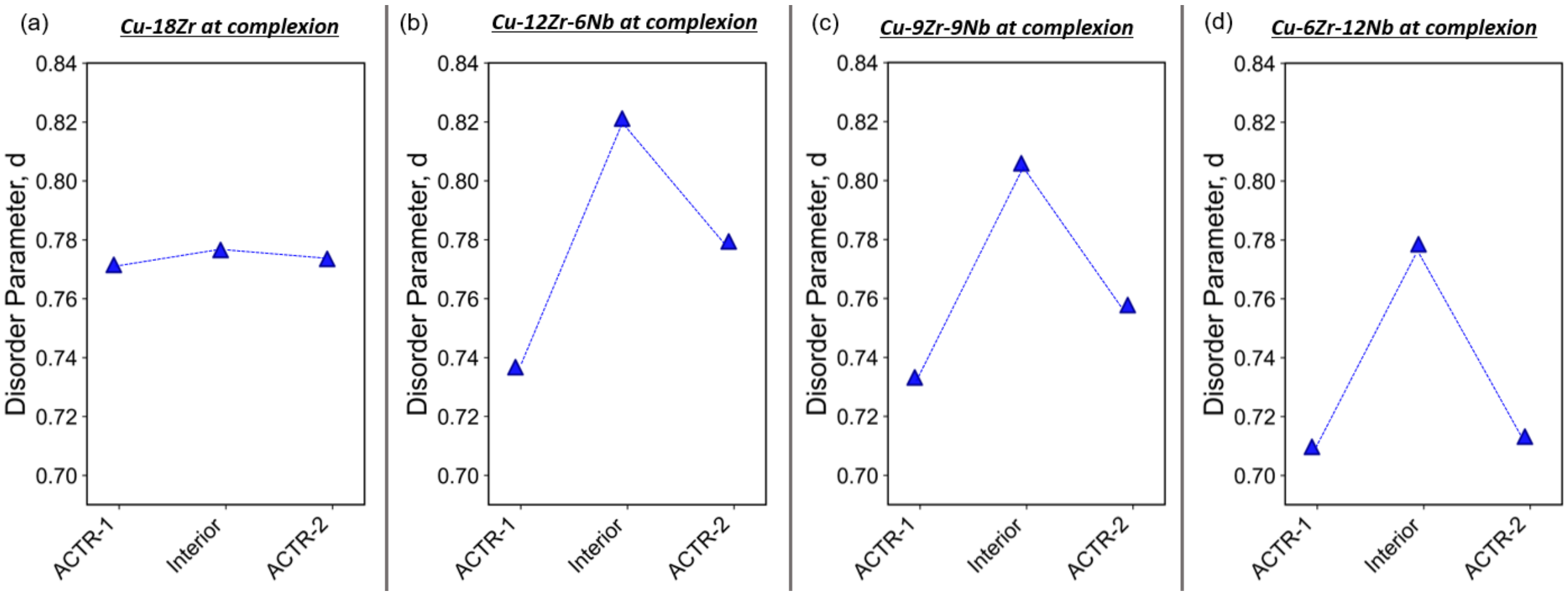

**Figure S19. Disorder parameters for different locations across an amorphous complexion with composition of (a) Cu-18Zr, (b) Cu-12Zr-6Nb, (c) Cu-9Zr-9Nb, and (d) Cu-6Zr-12Nb. The complexion interiors were more disordered than the ACTRs, and the level of asymmetry in disorder parameter for the ACTRs was dependent on the alloy concentration.**

As disorder parameter is calculated for each atom, this metric of local SSRO can be further examined in terms of each element, as shown in Figure S20. A number of overall trends can be found which extend across all alloys. First, the environment surrounding Cu atoms tends to be more ordered than that around the dopant species. The relative difference appears to be sensitive to the alloy system, ratio of dopant species, and the exact region of interest. For example, the elemental variations are generally larger in the ternary alloys than the binary alloy. In addition, ACTR-2 shows a much broader range of values for complexion with composition Cu-6Zr-12Nb (Figure S20(b)) as compared to the same region in Cu-12Zr-6Nb complexion (Figure S20(d)). Figure S21 provides a closer look of ACTR-2 in these two samples, presenting views of local disorder parameter within the ACTR plane. When comparing the Cu-12Zr-6Nb complexion (Figure S21(a)) to the Cu-6Zr-12Nb complexion (Figure S21(b), one can see that SSRO in the latter is significantly more heterogeneous. Figure S21(c) presents histograms of this same data, where the Cu-6Zr-12Nb complexion is found to have more sites with low disorder parameter (i.e., relatively ordered sites).

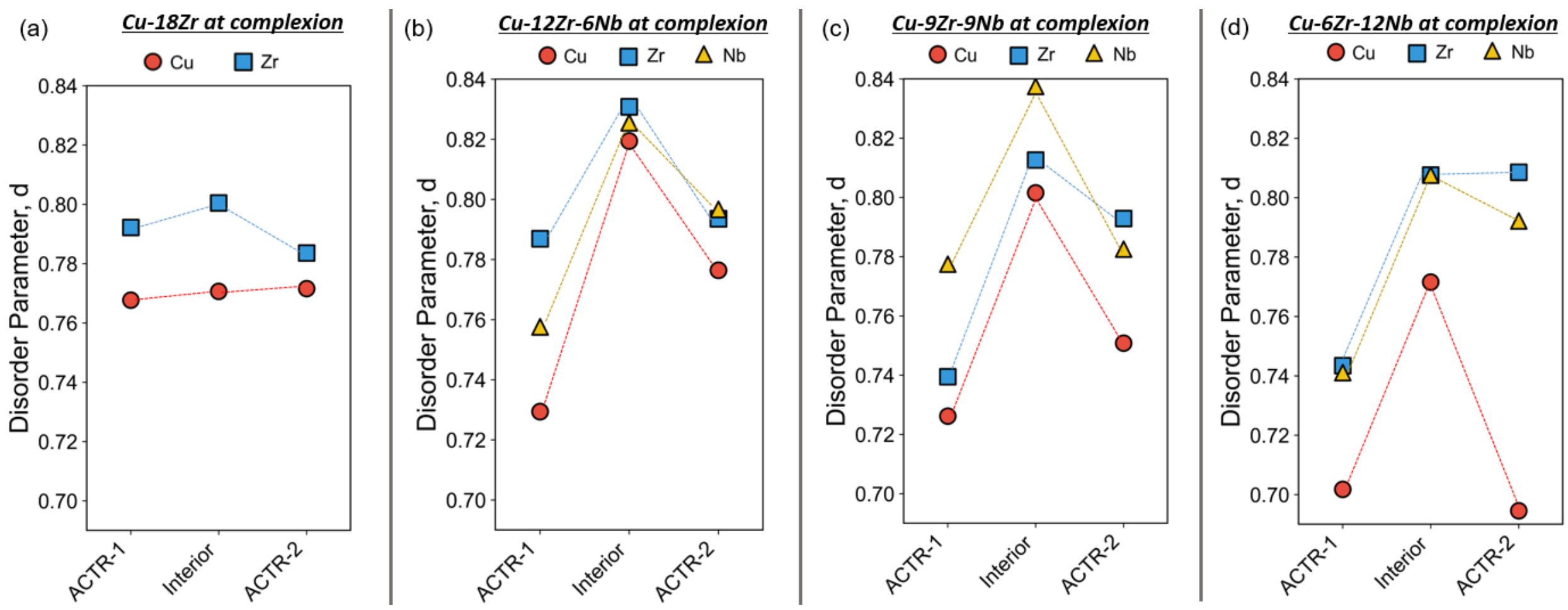

**Figure S20. Elemental disorder parameters for different locations across an amorphous complexion with composition of (a) Cu-18Zr, (b) Cu-12Zr-6Nb, (c) Cu-9Zr-9Nb, and (d) Cu-6Zr-12Nb. Cu atoms generally resided in more ordered environments than the dopant species, with the relative variations being alloy-dependent.**

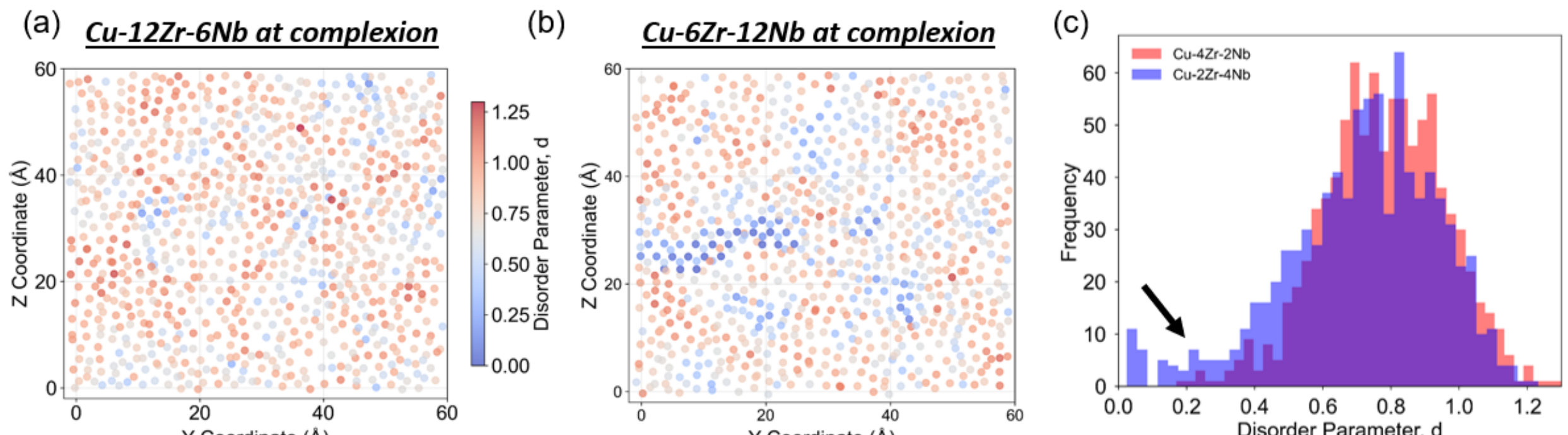

**Figure S21. In-plane distribution of the atomic disorder parameter for ACTR-2 in complexion (a) Cu-12Zr-6Nb and (b) Cu-6Zr-12Nb. (c) Histograms of the disorder parameter distribution for the two alloys, demonstrating that a collection of relatively ordered sites are present in Cu-2Zr-4Nb.**

**References.**